%% file: nuncs-arXiv.tex
\documentclass[twocolumn,showpacs,amsmath,prd,amssymb,showkeys,superscriptaddress]{revtex4}

\pdfoutput=1

\usepackage{multirow}
\usepackage{graphicx}
\usepackage{dcolumn}
\usepackage{bm}
\usepackage{amstext}
\usepackage{amsmath}
\usepackage{amssymb}
\usepackage{scalerel}

 \usepackage[unicode=true,pdfusetitle,
 bookmarks=true,bookmarksnumbered=false,bookmarksopen=false,
 breaklinks=false,pdfborder={0 0 1},backref=false,colorlinks=true, citecolor=blue]
           {hyperref}
           
\def\nuA{\nu {\rm A}_{el}}

\def\q2{q^2}
\def\Enu{{E_{\nu}}}
\def\T0{T_{min}}
\def\sigmanuA{\sigma_{\nuA}}

\def\nuebar{\bar{\nu}_e}

\def\hyphen{\mbox{-}}
\def\FF{F_{\rm A}}

\def\avealpha{\langle \alpha \rangle}

\def\avexi{\langle \xi \rangle}
\def\DARpi{\rm DAR \hyphen \pi}
\def\keVnr{\rm keV_{nr}}
\def\dRdT{\frac{ d R }{d T}}
\def\dRdq2{\frac{ d R }{d \q2 }}

\def\GNP{\Gamma_{\scaleto{NP}{4pt}} }
\def\GQM{\Gamma_{\scaleto{QM}{4pt}} }
\def\GDATA{\Gamma_{\scaleto{DATA}{3pt}} }

\begin{document}

\title{
Studies of Quantum-Mechanical
Coherency Effects in Neutrino-Nucleus Elastic Scattering
}

\newcommand{\as}{Institute of Physics, Academia Sinica,
Taipei 11529, Taiwan.}
\newcommand{\bhu}{Department of Physics, Institute of Science,
Banaras Hindu University,
Varanasi 221005, India.}
\newcommand{\cusb}{Department of Physics,
School of Physical and Chemical Sciences,
Central University of South Bihar, Gaya 824236, India}
\newcommand{\ntu}{
Department of Physics, CTS and LeCosPA, National Taiwan University,
Taipei 10617, Taiwan.}
\newcommand{\ndhu}{
Department of Physics, National Dong Hwa University,
Shoufeng, Hualien 97401, Taiwan.}
\newcommand{\deu}{Department of Physics,
Dokuz Eyl\"{u}l University, Buca, \.{I}zmir 35160, Turkey.}
\newcommand{\corr}{htwong@phys.sinica.edu.tw}

\author{ V.~Sharma }  \affiliation{ \as } \affiliation{ \bhu }
\author{ L.~Singh }  \affiliation{ \as } \affiliation{ \cusb }
\author{ H.T.~Wong } \altaffiliation[Corresponding Author: ]{ \corr } \affiliation{ \as }
\author{ M.~Agartioglu }  \affiliation{ \as } \affiliation{ \deu } \affiliation{ \ndhu }
\author{ J.-W.~Chen }  \affiliation{ \ntu }
\author{ M.~Deniz } \affiliation{ \deu }
\author{ S.~Kerman } \altaffiliation{Deceased} 
\affiliation{ \deu }
\author{ H.B~Li } \affiliation{ \as }  
\author{ C.-P.~Liu }  \affiliation{ \ndhu }
\author{ K.~Saraswat }  \affiliation{ \as } 
\author{ M.K.~Singh } \affiliation{ \as } \affiliation{ \bhu }
\author{ V.~Singh }   \affiliation{ \bhu } \affiliation{ \cusb }

\collaboration{ TEXONO Collaboration }

\date{\today}

\begin{abstract}

Neutrino-nucleus elastic scattering ($\nuA$)
provides a unique laboratory
to study the quantum-mechanical (QM) coherency effects in
electroweak interactions. 
The deviations of the cross-sections
from those of completely coherent systems
can be quantitatively characterized through
a coherency parameter $\alpha ( \q2 )$. 
The relations between $\alpha$ and 
the underlying nuclear physics
in terms of nuclear form factors are derived.
The dependence of cross-section
on $\alpha ( \q2 )$
for the various neutrino sources is presented.
The $\alpha ( \q2 )$-values are evaluated from
the measured data of the COHERENT CsI and Ar experiments.
Complete coherency and decoherency conditions 
are excluded by the CsI data  
with $p {=} 0.004$ at $\q2 {=} 3.1 {\times} 10^{3} ~ {\rm MeV^2}$
and
with $p {=} 0.016$ at $\q2 {=} 2.3 {\times} 10^{3} ~ {\rm MeV^2}$,
respectively,
verifying that both QM superpositions and
nuclear many-body effects contribute to $\nuA$ interactions.

\end{abstract}


\maketitle

\section{Introduction}
\label{intro}

The elastic scattering of a neutrino
with a nucleus~\cite{nuA-early,nuA-komas}
\begin{equation}
\nuA :
~~~~~~
\nu ~ + ~ A(Z,N) ~ \rightarrow ~
\nu ~ + ~ A(Z,N) ~~ ,
\label{eq::nuA}
\end{equation}
where $A(Z,N)$ denotes the atomic nucleus
with its respective atomic, charge and
neutron numbers,
is a fundamental electroweak neutral current process
in the Standard Model (SM).

Studies of neutrino-nucleus elastic scattering
can provide sensitive probes to
physics beyond SM (BSM)~\cite{nuA-BSM,scholberg}
and certain astrophysical processes~\cite{nuA-early,nuA-astro}.
It offers prospects to
study quantum-mechanical (QM) coherency effects in
electroweak interactions~\cite{PRD16},
neutron density distributions~\cite{nuA-nFF},
to detect supernova neutrinos~\cite{nuA-SN}
and to provide a compact and transportable
neutrino detectors for real-time
monitoring of nuclear reactors~\cite{nuA-monitor}.
The $\nuA$ events from
solar and atmospheric neutrinos are
the irreducible ``neutrino floor'' background~\cite{nuA-DMfloor}
to the coming generations of
dark matter experiments~\cite{rppdarkmatter}.

There are several active experimental programs
to observe and measure the $\nuA$ processes
with neutrinos from reactors~\cite{reactornuA} or
from decay-at-rest pions ($\DARpi$)~\cite{scholberg} provided by
spallation neutron source~\cite{SNSnuA}.
Future dark matter experiments may also
be sensitive to $\nuA$ from solar neutrinos~\cite{solarnuA}.
First positive measurement of $\nuA$
was achieved by 
the COHERENT experiment with 
CsI(Na) detector~\cite{COHERENT-CsI},
followed by measurements with 
liquid Ar detector~\cite{COHERENT-Ar}.

The $\nuA$ interaction provides a laboratory
to probe the QM coherency effects~\cite{PRD16}.
Experimental measurements are mostly performed in
a parameter space where the coherency effects are 
partial and incomplete. 
The deviations from perfect coherency
would have to be described and quantified before this
interaction can be effectively applied 
towards other goals like the studies of BSM physics.

Coherency in QM superpositions
among scattering amplitudes from 
individual nucleons is central to 
$\nuA$ interactions.
Our earlier work~\cite{PRD16} identified 
a coherency parameter $\alpha ( \q2 )$
which can quantify and consistently characterize
the degree of coherency
in $\nuA$ with different $\nu$-sources and target nuclei.
This article follows and expands on these studies.
The relations between $\alpha ( \q2 )$ with 
the complementary descriptions in terms of nuclear
physics via the language of nuclear form factors or
with the reduction in cross-sections
are discussed in Section~\ref{sect::basics}.
The dependence of coherency effects 
with interaction kinematics for various
neutrino sources and detector targets
are surveyed in Section~\ref{sect::configuration}.
The constraints provided by 
the COHERENT-CsI and -Ar data are
derived in Section~\ref{sect::coherent}.


\section{Formulation and Characterization}
\label{sect::basics}

The $\nuA$ differential cross-section 
at three-momentum transfer $q$ ($\equiv | \vec{q} |$) 
and neutrino energy $\Enu$
can be expressed as~\cite{nuA-komas,PRD16}:
\begin{equation}
\label{eq::dsigmadq2}
\left[ \frac{d \sigma}{d \q2} ( \q2 , \Enu ) \right] _{\nuA}   =   
\frac{1}{2} ~ \left[ \frac{ G_F^2 }{ 4 \pi}  \right] \cdot
\left[ 1 - \frac{ \q2 }{ 4 \Enu ^2 } \right] \cdot \Gamma ( \q2 )  ~ ,
\end{equation}
where $\Gamma ( \q2 )$ is a function describing
the contributions due to many-body physics in the target nuclei,
since the $\nuA$ interactions involve collective contributions 
of individual nucleons in the nucleus. 

The relevant kinematics variable is $\q2$
which characterizes the physics
and is universal to all target.
The experimental observable is the
nuclear recoil energy ($T$),
expressed in units of $\keVnr$ in this article,
which depends on the target nuclear mass $M$
and is related to $\q2$ via
$ \q2 {=} 2 M T {+} T^2 {\simeq} 2 M T $.
The minimal observable energy $\T0$ for the nuclear recoils is
the detector threshold, while
kinematics limits the maximum
recoil energy to be
$T_{max} {=} 2 \Enu ^2 / ( M  {+} 2 \Enu ) {\simeq} 2 \Enu ^2 / M$.
These limits can be translated to
$\q2_{min}$=$2 M \T0$ and
$\q2_{max}$=$4 \Enu ^2 [ M / ( M$+$2 \Enu ) ]$$\simeq$$4 \Enu ^2$.
The variations of the 
$\nuA$ differential and integral cross-sections 
with respect to $T$ are discussed in
Appendix~\ref{app::rates}.

Depending on the particular physics aspects to probe,  
there are complementary formulations on the $\Gamma ( \q2 )$ function. 
The conventional description is based on nuclear physics, 
expressed as 
\begin{equation}
\Gamma ( \q2 ) \equiv \GNP ( \q2 ) =
 \left[ \varepsilon Z F_Z ( \q2 ) - N F_N ( \q2 ) \right] ^2 ~~ ,
\label{eq::GNP}
\end{equation}
where $F_Z ( \q2 ) {\in} [0,1]$ 
and $F_N ( \q2 ) {\in} [0,1]$ are, respectively,
the proton and neutron nuclear form factors
for the nucleus $A(Z,N)$, while
$\varepsilon$$\equiv$(1$- 4 ~ {\rm sin^2 \theta_W } )$=0.045,
indicating the dominant contributions are
from the neutrons.

The merit of this description is to connect $\nuA$ to 
nuclear physics so that its studies may benefit from
or contribute to the wealth of information and data.
Electron-nucleus scattering experiments provide
important data to the
nuclear proton form factor $F_Z ( \q2 )$~\cite{protonFF}.
The neutron counterpart $F_N ( \q2 )$, however,
would require weak processes to probe.
Studies of $\nuA$ have therefore triggered intense activities
towards their measurements~\cite{nuA-nFF},
complementing experiments with
parity-violation scattering
using polarized electrons~\cite{PVeN}.


\begin{figure}
{\bf (a)}\\
\includegraphics[width=8.2cm]{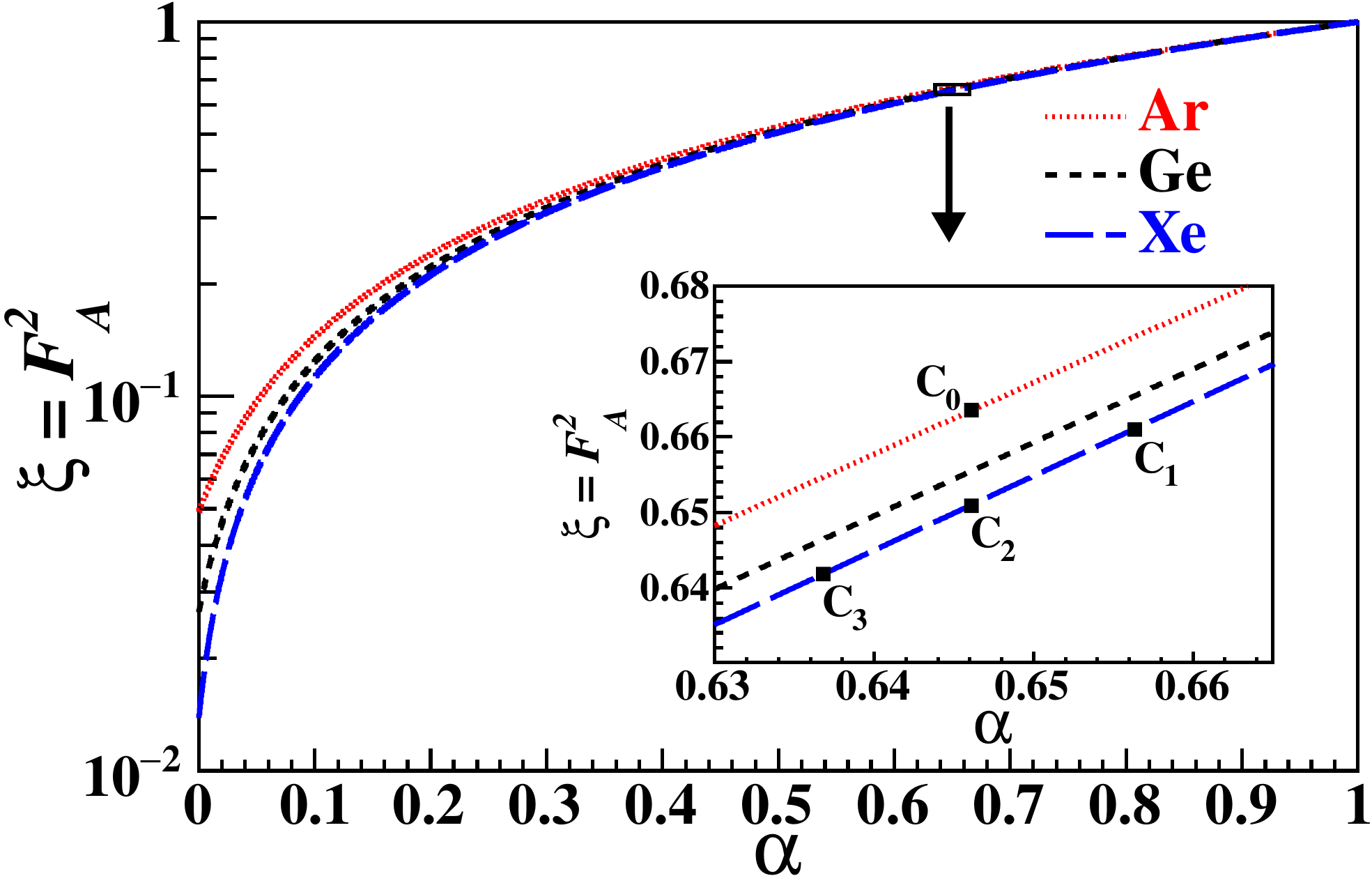}\\
{\bf (b)}\\
\includegraphics[width=8.2cm]{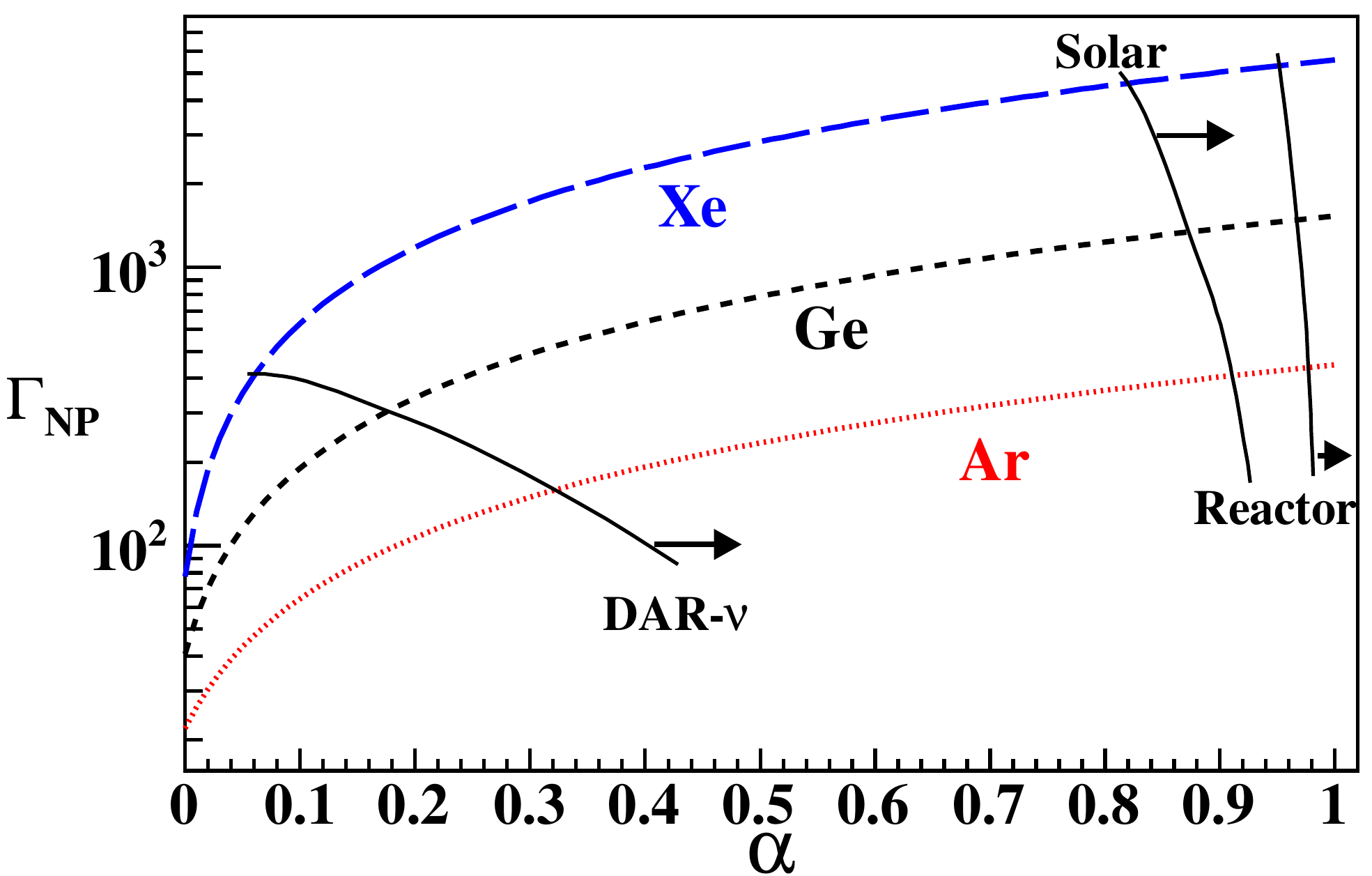}
\caption{
The variations with $\alpha$ for
(a) cross-section reduction fraction $\xi$ (equivalently $\FF$) and
(b) $\GNP$
for the three target nuclei, independent of
underlying nuclear physics.
Superimposed in (b) are contours of maximum-$\q2$ 
for reactor, solar and $\DARpi$ neutrinos.
The limiting values are 
$\xi$=1 and $\GNP {=} ( \varepsilon Z {-} N )^2$ 
at $\alpha$=1 and
$\xi {=} ( \varepsilon ^2 Z {+} N )/( \varepsilon Z {-} N )^2$ 
and $\GNP {=} ( \varepsilon ^2 Z {+} N )$ at $\alpha$=0.
The configurations ${\rm C_{0,1,2,3}}$ 
in the inset of (a) illustrate the cases 
where $\alpha_0 {<} \alpha_1$,  
$\alpha_0 {=} \alpha_2$ and 
$\alpha_0 {>} \alpha_3$ despite having 
$F_A^2 {=} \xi$ and 
$\xi_0 {>} \xi_{1,2,3}$
in all cases. 
}
\label{fig::ParaVsalpha}
\end{figure}


\input{nuncs-Table1}


In the kinematics regime relevant to this work $-$
$ \q2 R ^2 {\ll} \pi ^2$ 
(natural units with $\hbar {=} c {=} 1$ are used throughout),
where $R {=} 1.2 A^{1/3} ~ {\rm fm}$ is the typical 
scale characterizing the radius of nuclei $-$
nucleons can be taken as structureless point-like particles,
such that their internal dynamics and QCD effects
can be neglected.
At $\q2 {\rightarrow} 0$, 
there is a perfect alignment of
the scattering amplitude vectors of 
individual nucleons in the target nucleus~\cite{PRD16}.
The interactions are completely coherent.
As $\q2$ increases, deviation from 
this complete coherency condition
leads to reduction in the cross-section.
The degree of coherency can be quantified
by a parameter 
$\alpha ( \q2 ) {\equiv} {\rm cos }  { \phi } {\in} [0,1]$
where ${ \phi ( \q2 )} {\in} [0,\pi/2]$ is 
the misalignment phase angle
between  the scattering amplitudes of two non-identical
nucleons~\cite{PRD16}.
This leads to a formulation in terms of 
QM superpositions among the
various scattering centers, in which:
\begin{eqnarray}
\Gamma ( \q2 ) & \equiv & \GQM ( \q2 ) \\
& = &   Z\varepsilon^{2} 
\left[1 {+} \alpha(Z{-}1) \right] {+} N \left[1 {+} \alpha(N{-}1) \right]  
{-} 2\alpha\varepsilon Z N   \nonumber \\
& = &   ( \varepsilon Z - N ) ^2 \cdot \alpha ( \q2 ) ~  + ~ 
( \varepsilon ^2  Z + N ) \cdot 
\left[ 1 - \alpha ( \q2 ) \right] ~ .  \nonumber
\label{eq::alpha}
\end{eqnarray}

The $\GQM$-formulation with $\alpha ( \q2 )$ 
provides an intuitive physics understanding 
and quantitative description on the 
reduction of $\nuA$ cross-sections in terms of 
QM phase-angle alignment and coherency.
In particular, it naturally leads to
the limiting behavior at the
complete coherency ($\alpha {=} 1$ at $\q2 {\sim} 0$)
and
decoherency ($\alpha {=} 0$ at $\q2 {\gtrsim} [ \pi / R ]^2$) 
states, 
corresponding to $( d\sigma/d\q2 ) {\propto} [ \varepsilon Z - N  ]^2 $  and
$( d\sigma/d\q2 ) {\propto} [\varepsilon^2 Z + N] $, respectively.
The experimentally measured $\alpha ( \q2 )$-values 
from different isotope targets can be directly compared 
to reveal their varying degrees of coherency 
in the respective processes.

An alternative measurement-driven description,
denoted by $\xi ( \q2 )$,  
is the cross-section reduction 
relative to that of complete coherency condition~\cite{PRD16}, where
\begin{equation}
\Gamma ( \q2 )  \equiv  \GDATA ( \q2 ) 
 = ( \varepsilon Z - N ) ^2  \cdot \xi ( \q2 ) ~~ .  
\end{equation}

The functions $\GNP$, $\GQM$ and $\GDATA$ are complementary
descriptions of the $\nuA$ interactions.
The experimentally measurable 
cross-section reduction fraction
($\xi$ in $\GDATA$) 
is related to
QM coherency ($\alpha$ in $\GQM$)
and
nuclear form factors 
via, respectively,
\begin{equation}
\xi ( \q2 ) 
 =  \alpha ( \q2 )+
\left[ 1 - \alpha ( \q2 ) \right]
\left[   \frac{ ( \varepsilon ^2  Z + N ) }
{ ( \varepsilon  Z - N  ) ^2 } \right]
\label{eq::xiVsalpha}
\end{equation}
and
\begin{equation}
\xi ( \q2 ) =
\frac{\left[ \varepsilon Z F_Z ( \q2 ) -  N F_N ( \q2 ) \right] ^2} 
{ ( \varepsilon  Z - N  ) ^2 } ~ ~ ,
\label{eq::xiVsGNP}
\end{equation}
while
the two physics descriptions are connected by:
\begin{eqnarray}
\label{eq::FFVsalpha}
\left[ \varepsilon Z F_Z ( \q2 ) -  N F_N ( \q2 ) \right] ^2 
& = &
( \varepsilon Z - N ) ^2 \cdot \alpha ( \q2 ) ~  + \\
& & 
( \varepsilon ^2  Z + N ) \cdot 
\left[ 1 - \alpha ( \q2 ) \right] ~ . \nonumber
\end{eqnarray}

The relations between $\xi$ and $\GNP$ with $\alpha$ for
three representative 
nuclei are shown in Figures~\ref{fig::ParaVsalpha}a\&b,
respectively.
Contours of maximum-$\q2$ for different neutrino sources are 
marked in Figure~\ref{fig::ParaVsalpha}b.
The behavior of $\GNP$, $\alpha$ and $\xi$
at the limiting domains 
corresponding to the complete coherency and 
decoherency conditions
are summarized in Table~\ref{tab::QMcoherency}.
In particular, the relation 
$\GNP {=} ( \varepsilon ^2  Z + N )$ 
for completely decoherent $\nuA$ interactions
is a result that emerges by relating $\GNP$ and $\GQM$
in Eq.~\ref{eq::FFVsalpha},
and could not be derived by considerations of 
nuclear form factor of Eq.~\ref{eq::GNP} alone.


\section{Projected Experimental Ranges}
\label{sect::configuration}

The functions $\GNP$, $\GQM$ and $\GDATA$ 
can be directly measured from $\nuA$ data
without input from the underlying physics.
Prior to actual measurements,
specific formulations of the
nuclear form factors have to be adopted
for phenomenological studies and
to establish the typical ranges 
to guide the choices 
of experimental parameters.
To serve these purposes, 
the frequently adopted approach is to take
the nuclear form factors for protons and neutrons are identical: 
$ F_N ( \q2 ) {=} F_Z ( \q2 ) {\equiv} \FF ( \q2 )$,
and to use the effective 
``Helm Form Factor'' description of Ref.~\cite{engel}:
\begin{equation}
\FF ( \q2 ) =
\left[ \frac{3}{q R_0} \right] ~
j_1 ( q R_0 )
~ {\rm exp}  \left[ - \frac{1}{2} \q2 s^2 \right]  ~~~ ,
\label{eq::formfactor}
\end{equation}
where 
$j_1 ( x ) {=} [ ( {\rm sin} x / x^2 ) {-} ({\rm cos} x / x ) ]$ is
the first-order spherical Bessel function.
The nuclear dependence appears through
$R_0^2 {=} R^2 {-} 5 s^2$,
where $s {=} 0.5~{\rm fm}$ 
is the surface thickness of the nuclei.
In this formulation, the squared-form factor 
is equivalent to the cross-section reduction fraction:
$[ \FF ( \q2 ) ]^2 {=} \xi ( \q2 )$.


\begin{figure}[t] 
\includegraphics[width=8.2cm]{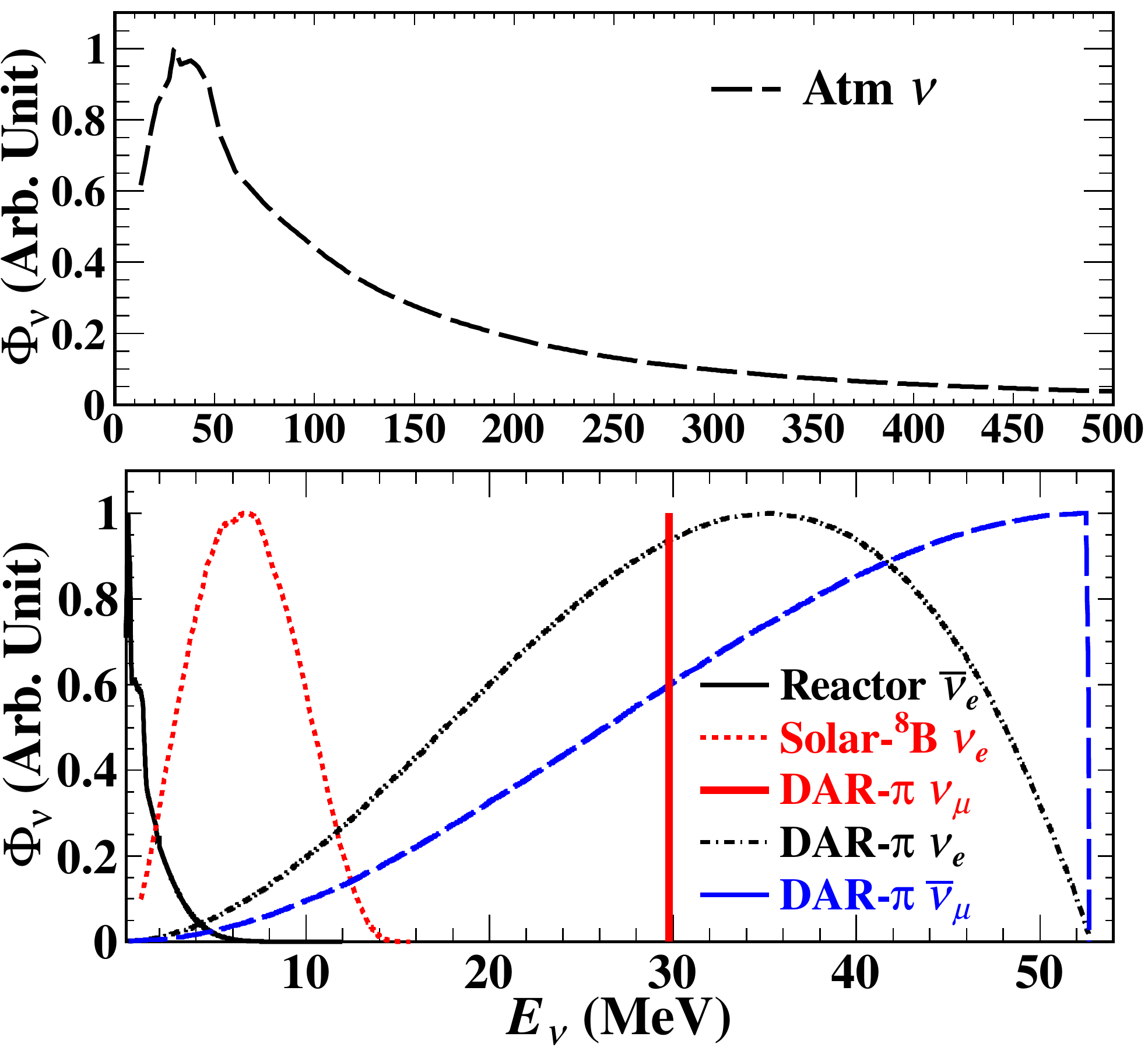}
 \caption{
Neutrino spectra ($\Phi_{\nu}$) from 
(top) atmospheric as well as (bottom)
reactor $\bar{\nu}_{e}$,
solar $^8$B $\nu_{e}$  and
$\DARpi$ ($\nu_{\mu}$, $\nu_{e}$, $\bar{\nu}_{\mu}$) neutrinos
adopted from Refs.~\cite{COHERENT-CsI,nuspectra} and normalized by their maxima.
}
\label{fig::nuspectra}
\end{figure}


\begin{figure}
\includegraphics[width=8.2cm]{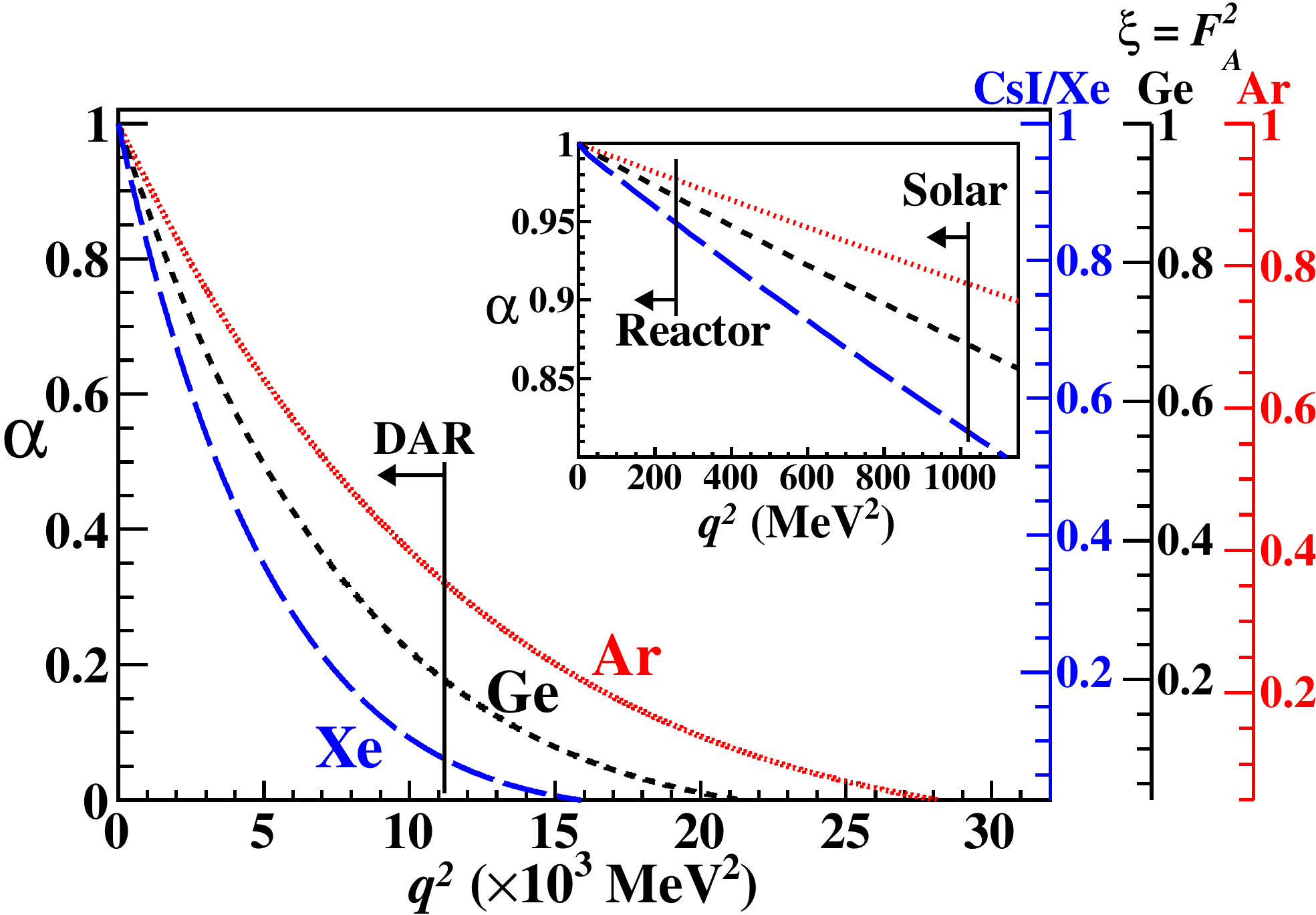}\\
\caption{
The variation of
$\alpha$ and $\xi$($=\FF^2$)
as a function of $\q2$
of $\nuA$ on the three selected nuclei.
Different neutrino sources share the same
contour for the same target in $\q2$-space,
but with different ranges. 
The end-points for reactor, solar and $\DARpi$ neutrinos
are marked.
}
\label{fig::q2-dependence}
\end{figure}


\begin{figure}
\includegraphics[width=8.2cm]{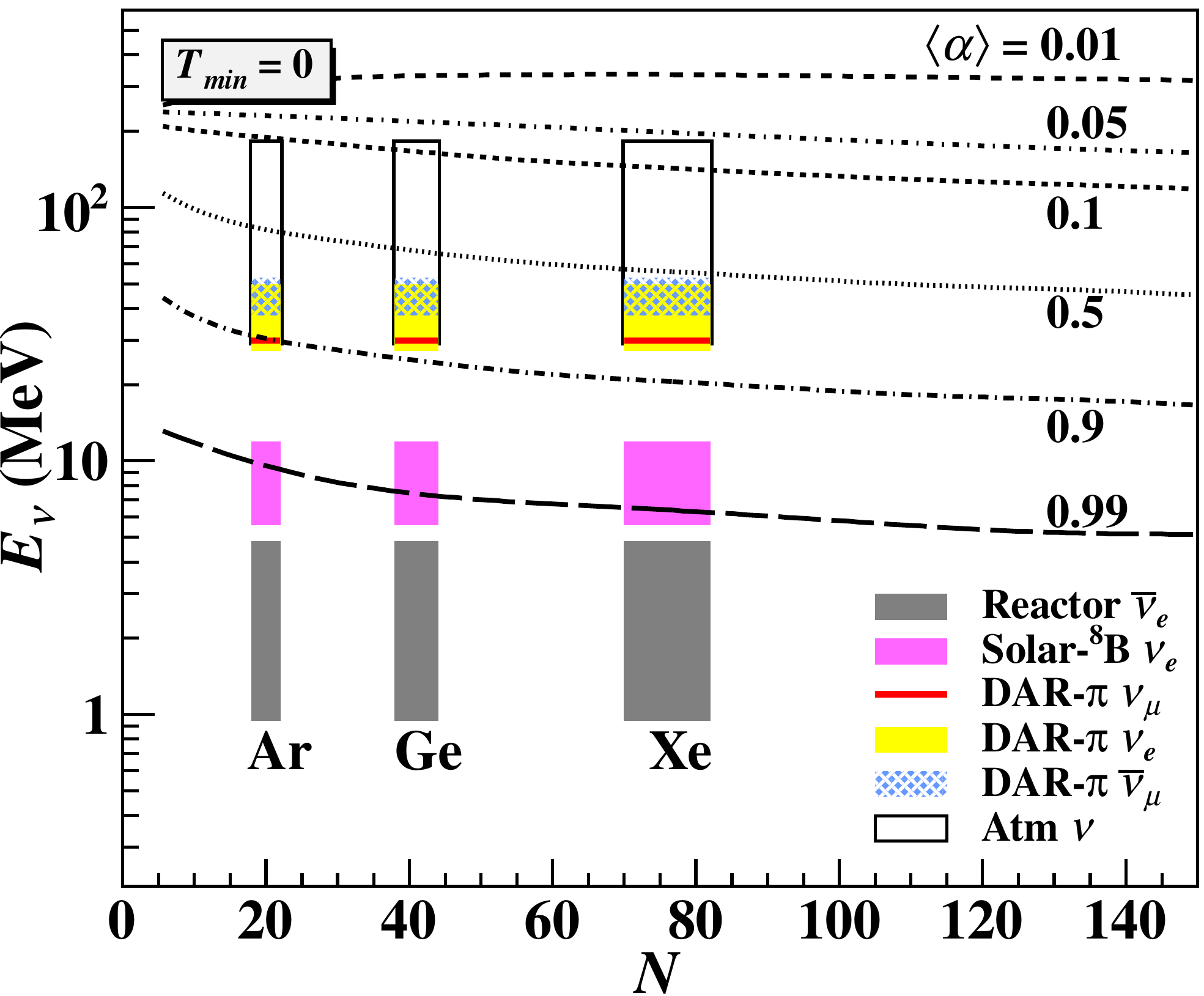}
 \caption{
The contours of the mean degree of coherency $\avealpha$
on the ($N$,$\Enu$) plane at $\T0$=0,
with bands of neutrino sources and target nuclei superimposed.
The ranges in $\Enu$ correspond to FWHM in
$[ \Phi_{\nu} {\cdot} \sigmanuA ]$.
}
\label{fig::summary}
\end{figure}


Typical spectra
of reactor, solar and atmospheric neutrinos~\cite{nuspectra},
as well as those due to
decay-at-rest $\pi$ ($\DARpi$)~\cite{COHERENT-CsI}, 
are used in this study. These are 
depicted in Figure~\ref{fig::nuspectra}.

The measurable total cross-section is given by
convoluting Eq.~\ref{eq::dsigmadq2}
with the neutrino spectrum $\Phi_{\nu} ( \Enu )$,
and integrating over
$\Enu$ and $\q2 {\in} [ \q2_{min} , \q2_{max} ]$,
from which the mean reduction fraction $\avexi$
and the mean coherency factor $\avealpha$ can be
derived~\cite{footnote1}.

The $\nuA$ processes on several nuclei 
of experimental interest 
and at different mass ranges 
are studied
$-$ (Ar;Ge;Xe) with $Z$=(18;32;54). 
The target that provides the first 
$\nuA$ measurements~\cite{COHERENT-CsI} $-$ CsI, 
having $Z$=55 and 53, respectively,
can be approximated as Xe in this discussion.


The variations of
$\alpha$ and $\xi$(=$\FF^2$)  
with $\q2$ of the four 
neutrino sources, with three selected nuclei (Ar;Ge;Xe) 
are depicted in Figure~\ref{fig::q2-dependence}. 
The $\q2$-dependence is universal  
for the different neutrino sources,
though their $\q2_{max}$-values are distinct  
due to their varying maximum $\Enu$.
These spectra end-points for reactor, solar and $\DARpi$ neutrinos
are well-defined, and their 
corresponding ranges in $\alpha$ and $\q2$
are depicted in 
Figures~\ref{fig::ParaVsalpha}b\&\ref{fig::q2-dependence}, respectively. 

A summary plot on the variations of $\avealpha$ 
with the neutrino sources and target nuclei is 
illustrated in Figure~\ref{fig::summary},
in which the ranges in $\Enu$ are defined by
the Full-Width-Half-Maximum (FWHM) of
$[ \Phi_{\nu} {\cdot} \sigmanuA ]$.
For completeness,
the differential and integral
event rates due to the four neutrino sources
in measurable nuclear recoil energy $T$,
together with their corresponding $\alpha$ 
and $\avealpha$ values,
are discussed and presented in Appendix~\ref{app::rates}. 

It can be seen that 
coherency is mostly complete ($\alpha {>} 95\%$)
for $\nuA$ with reactor and solar neutrinos, 
whereas coherency is only partial for $\DARpi$
and weak for atmospheric neutrinos.
Accordingly, studies of $\nuA$ with different neutrino sources
provide complementary information and cover the
transitions from completely coherent to decoherent states.


\section{Measurements from Current Data}
\label{sect::coherent}

The COHERENT-CsI(Na) and -Ar experiments at
the $\DARpi$ beam with the Spallation Neutron
Source facility at the Oak Ridge National Laboratory
have provided positive measurements on $\nuA$.

While the first-generation `discovery''
measurements cannot be expected
to provide severe constraints on $\alpha ( \q2 )$,
it is instructive to go through the data analysis
to establish the ranges of the effects
and to check consistency.


\begin{figure}
\hspace*{0cm}
{\bf (a)}\\
\includegraphics[width=8.2cm]{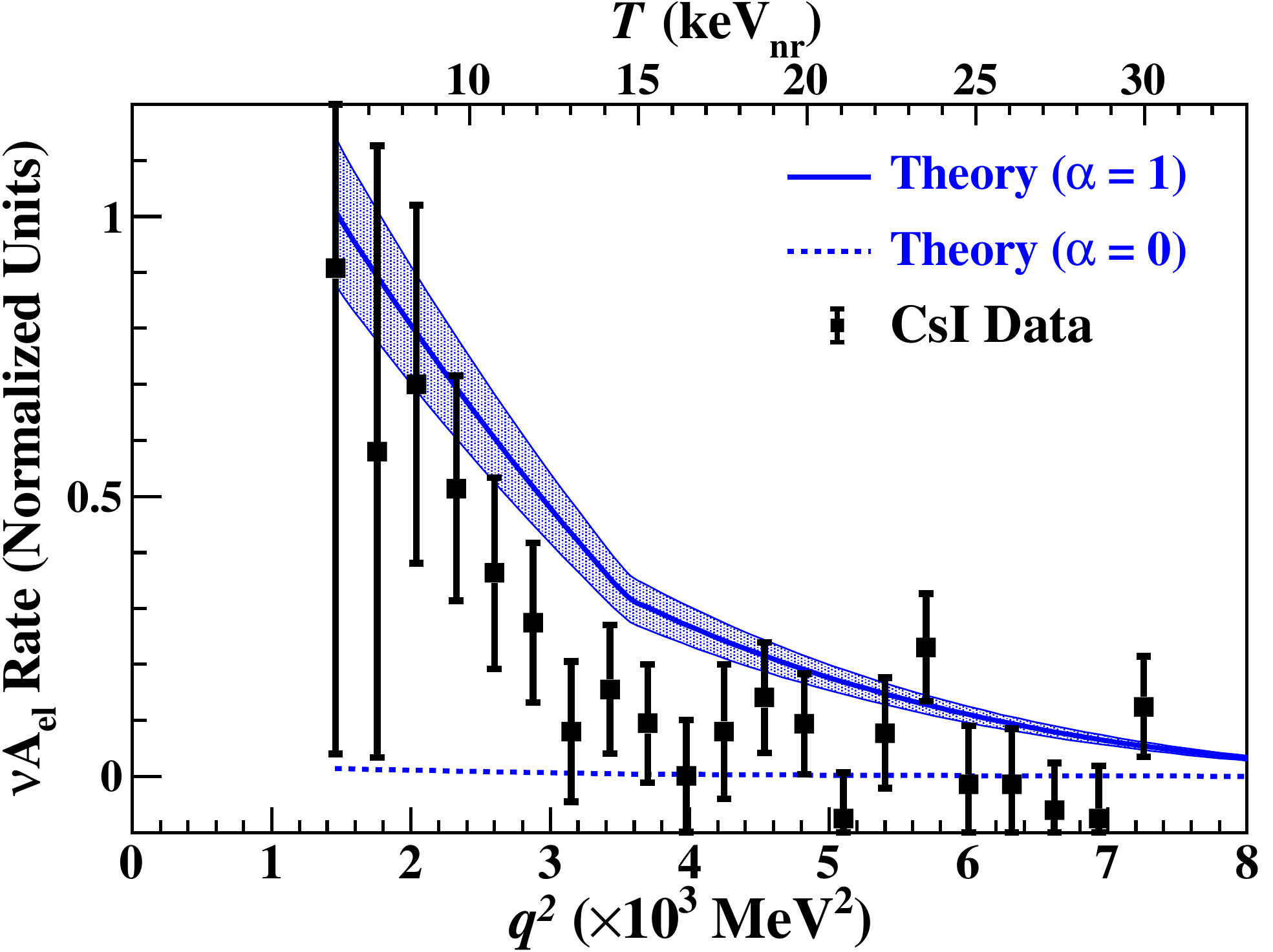}\\
{\bf (b)}\\
\includegraphics[width=8.2cm]{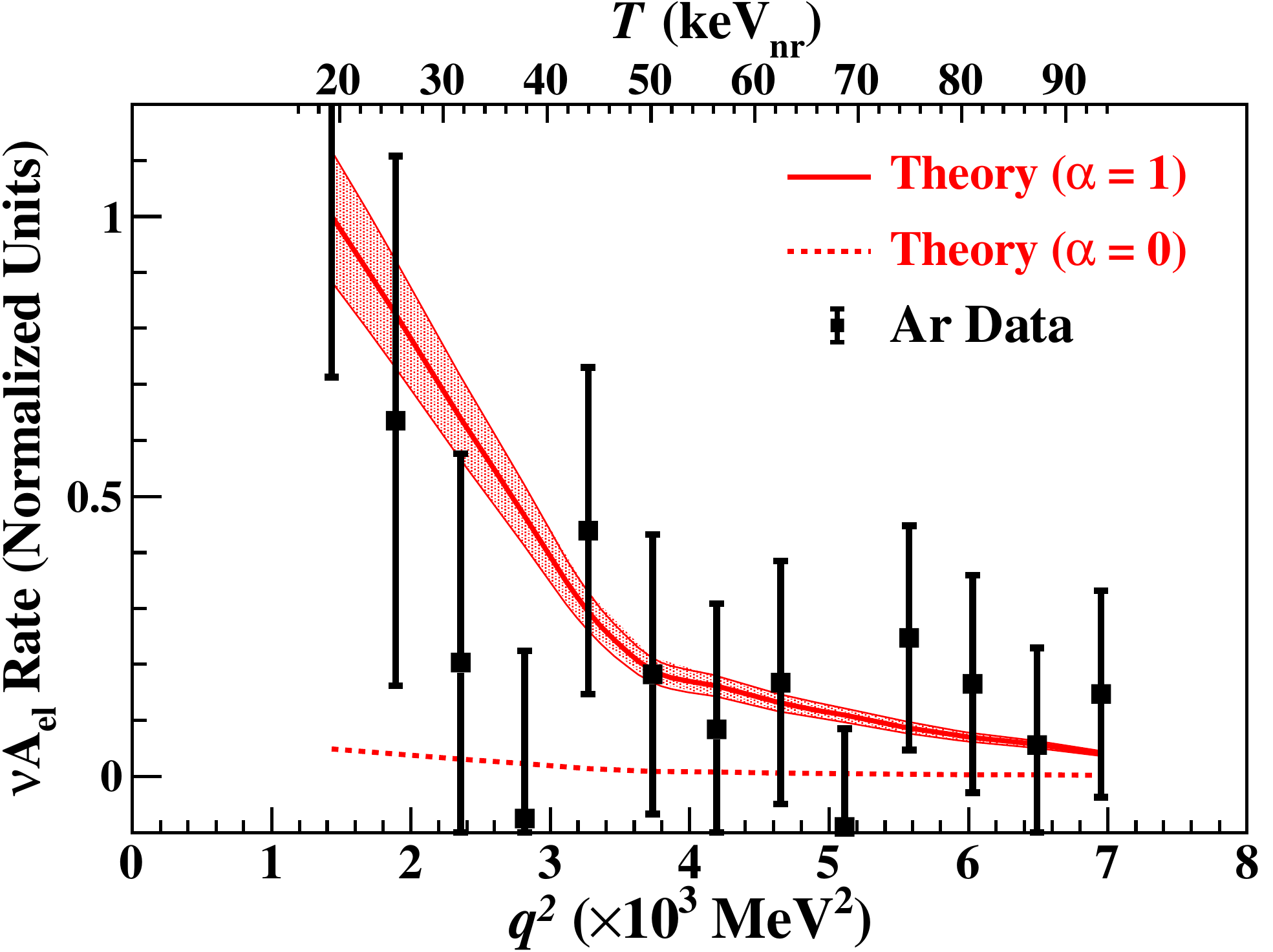}
\caption{
Efficiency-corrected differential event rates 
with their statistical uncertainties
derived from the COHERENT 
(a) CsI(Na)~\cite{COHERENT-CsI,COHERENT-QF-2020} and 
(b) Ar~\cite{COHERENT-Ar} data.
Superimposed are the predicted theory bands
at the complete coherency ($\alpha {=} 1$) condition, where 
$\Gamma ( \q2 ) {=}  ( \epsilon Z {-} N )^2$ 
is set in Eq.~\ref{eq::dsigmadq2}.
Systematic uncertainties are represented by
the width of the bands.
Their ratios give rise to
the cross-section reduction ratios $\xi$,
from which $\alpha ( \q2 )$ is derived 
via Eq.~\ref{eq::xiVsalpha}.
The allowed intervals and p-values
follows from standard Gaussian statistics~\cite{rppstatistics}.
The maxima of the Theory bands within the analyzed ranges 
are normalized to unity.
The complete decoherency conditions ($\alpha {=} 0$)
are denoted by dotted lines.
}
\label{fig::eff_correct}
\end{figure}

\begin{figure}
\hspace*{0cm}
{\bf (a)}\\
\includegraphics[width=8.2cm]{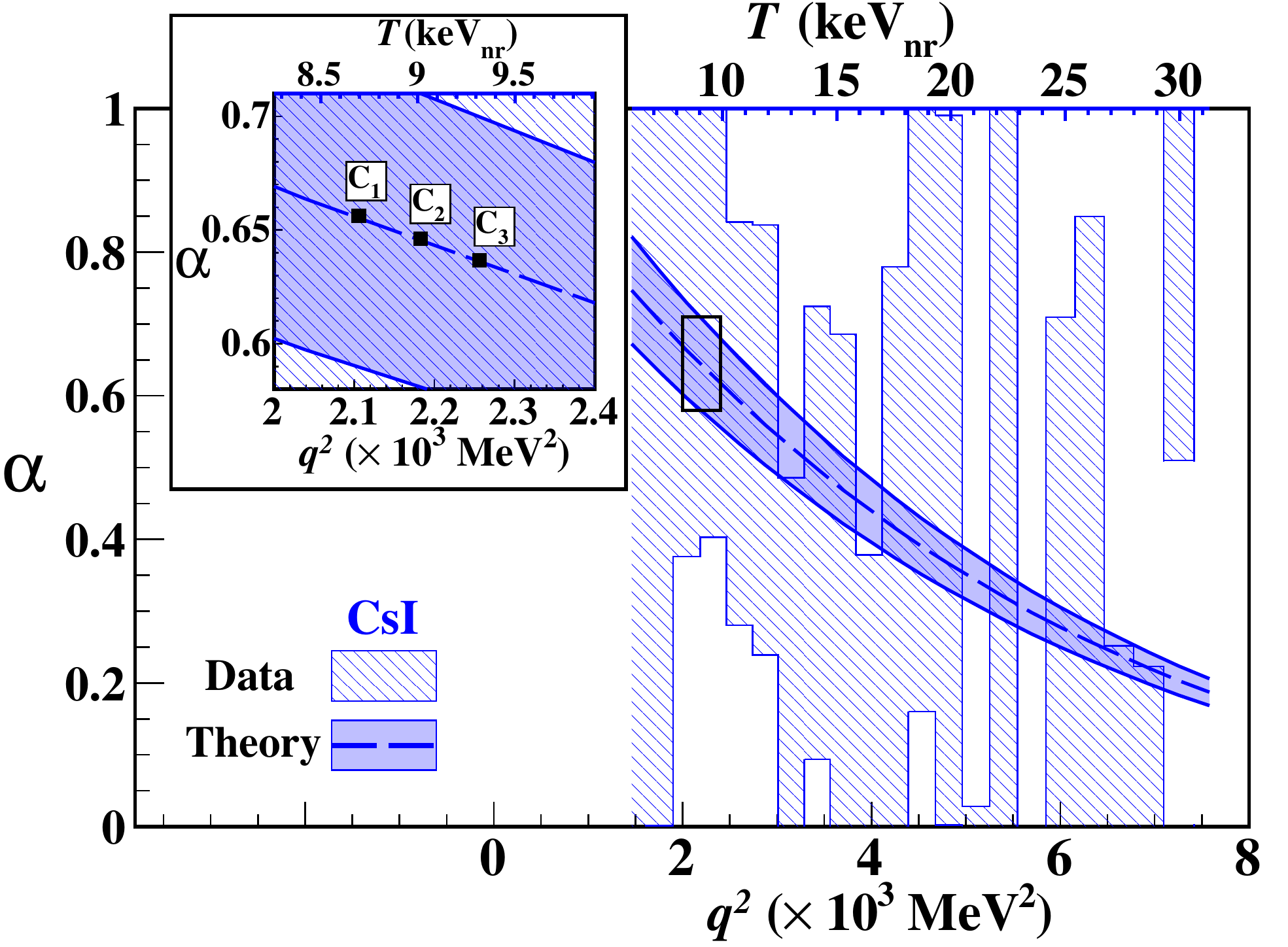}\\
{\bf (b)}\\
\includegraphics[width=8.2cm]{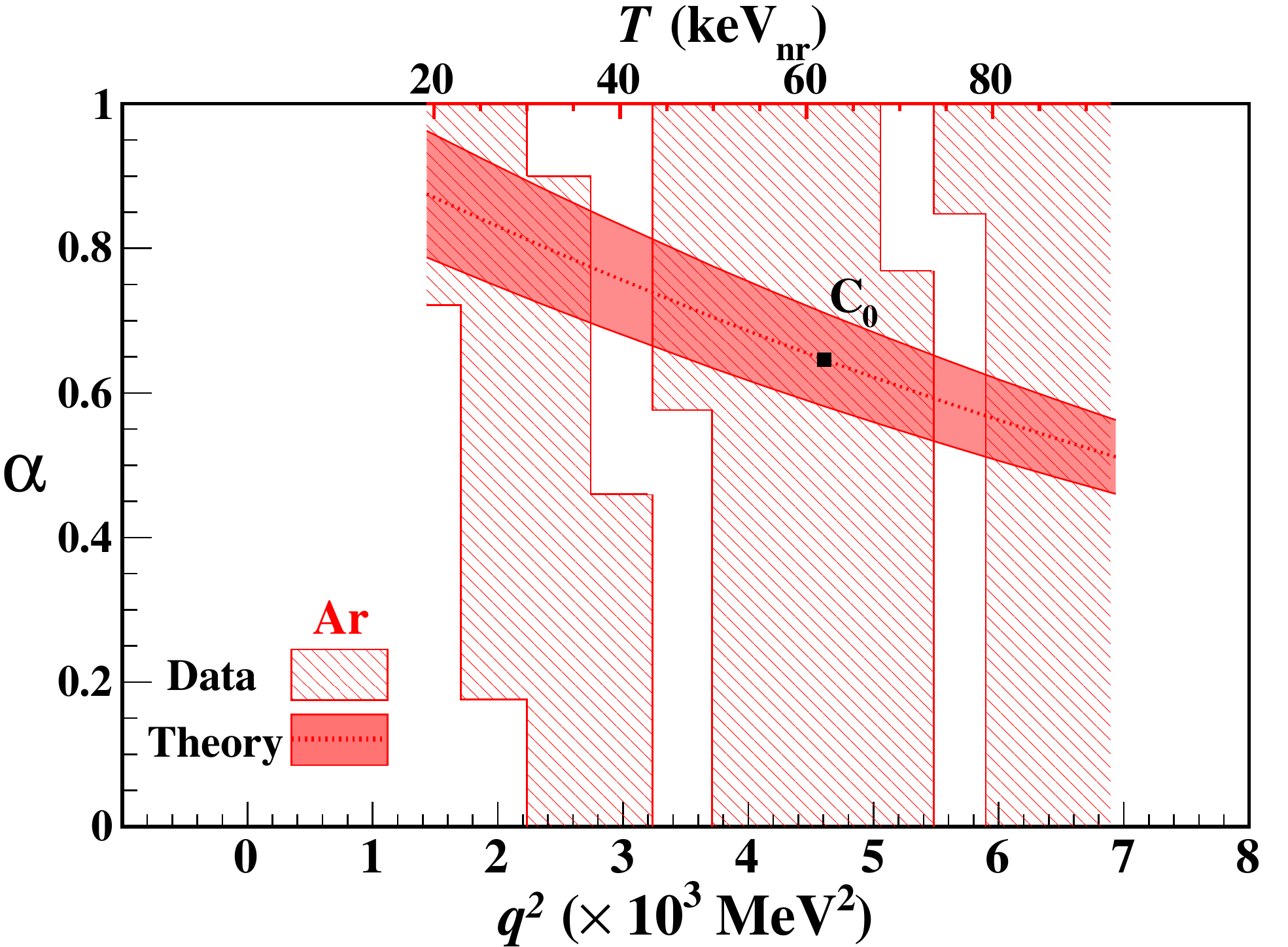}\\
{\bf (c)}\\
\includegraphics[width=8.2cm]{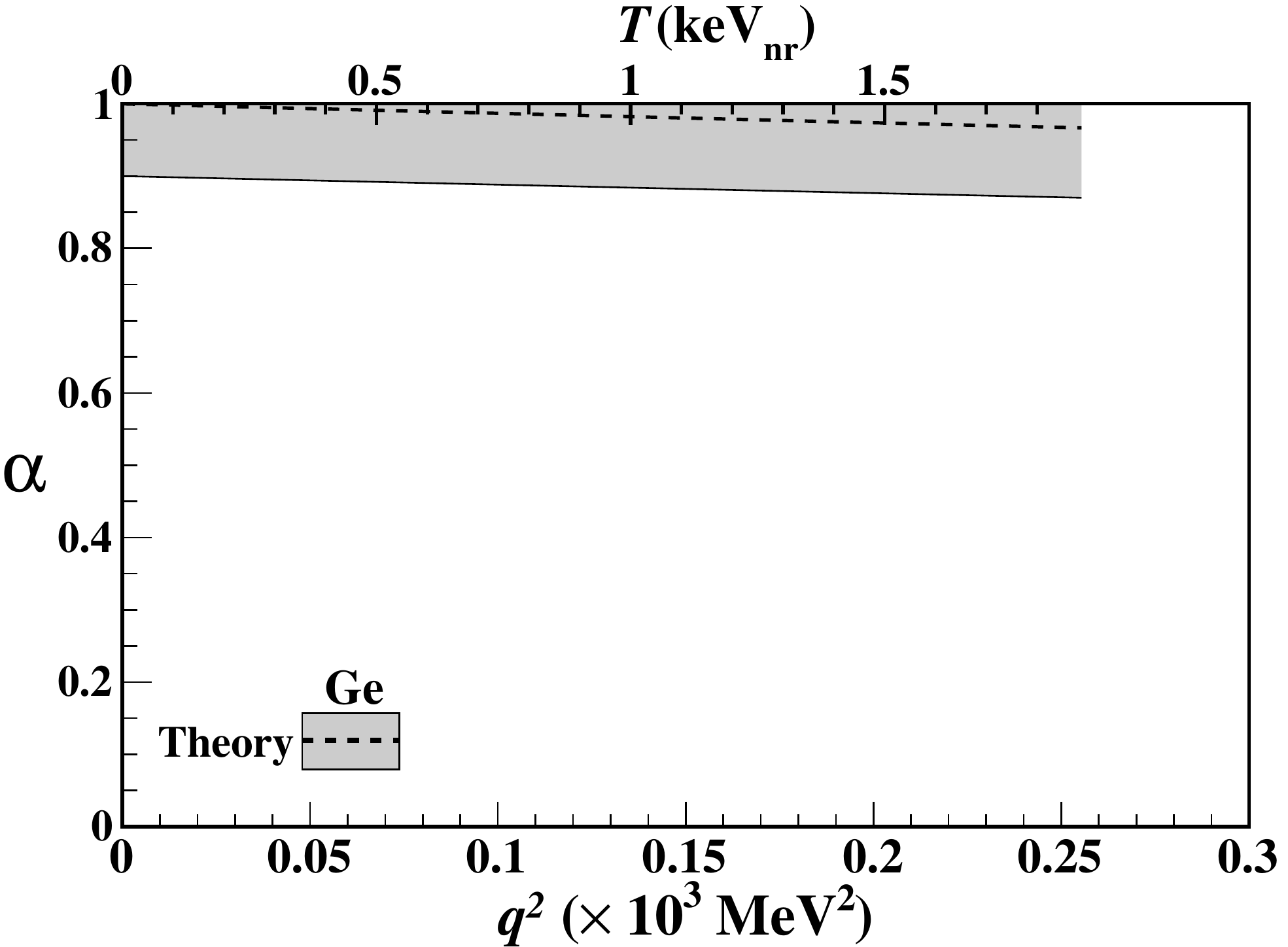}\\
\caption{
Measurements on $\alpha$ from 
COHERENT (a) CsI~\cite{COHERENT-CsI,COHERENT-QF-2020}
and (b) Ar~\cite{COHERENT-Ar} data 
with $\DARpi$-$\nu$. 
The stripe-shaded areas 
are the 1-$\sigma$ allowed regions
derived from the reduction in cross-section 
relative to the complete coherency conditions
independent of nuclear physics input.
The dark-shaded regions are 
the theoretical expectations 
adopting the nuclear form factor formulation of
Eq.~\ref{eq::formfactor} with a 
$\pm 1 \sigma$ uncertainty of 10\%. 
The $\alpha ( \q2 )$-values for 
different nuclei can be consistently compared.
Labels ${\rm C_{0,1,2,3}}$ 
correspond to the
configurations in Figure~\ref{fig::ParaVsalpha}a
where $\alpha_0 {<} \alpha_1$,  
$\alpha_0 {=} \alpha_2$ and 
$\alpha_0 {>} \alpha_3$ 
despite having $\xi_0 {>} \xi_{1,2,3}$
in all cases. 
(c) The sensitivity
with the theoretical projections applied
to reactor-$\nu$ on Ge 
covering the complete
$\q2$-range for nuclear recoils.
}
\label{fig::alpha-constraints}
\end{figure}

\begin{figure}
\hspace*{0cm}
\includegraphics[width=8.2cm]{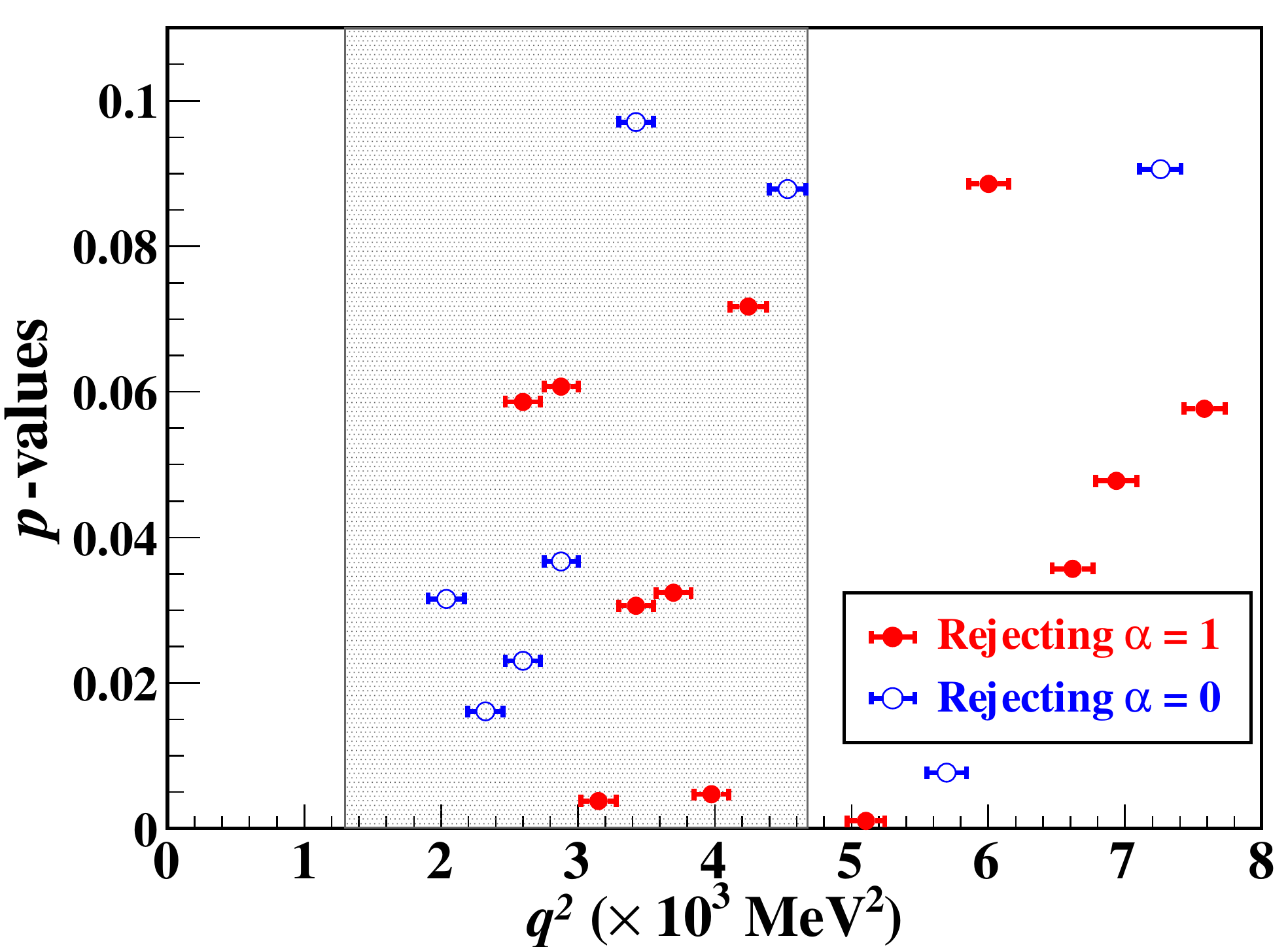}
\caption{
The $p$-value significance to
probe the specific cases corresponding to the
complete coherency ($\alpha$=1 and $\phi$=0, in red)
and decoherency($\alpha$=0 and $\phi {=} \pi / 2$, in blue) conditions
from the COHERENT-CsI data~\cite{COHERENT-CsI,COHERENT-QF-2020}.
The shaded band corresponds to the stated region-of-interest
in Ref.~\cite{COHERENT-CsI}
where physics analysis was performed.
}
\label{fig::p-values}
\end{figure}


The published event rates and statistical uncertainties
from the COHERENT CsI(Na)~\cite{COHERENT-CsI} 
and Ar~\cite{COHERENT-Ar} data 
were adopted as input in this analysis. 
What were measured per event are the
numbers of photo-electrons.
To convert these to 
nuclear recoil energy for physics interpretation,
the knowledge of quenching factor (QF) is necessary.
The Ar results of Ref.~\cite{COHERENT-Ar} has QF
incorporated already.
The uncertainties of the QF-model used to derive 
the first results of CsI(Na) in Ref.~\cite{COHERENT-CsI} 
were at $\sim$25\%.
For this analysis, a
subsequent improved QF-measurement
by the same Collaboration~\cite{COHERENT-QF-2020}
were adopted,
in which an accuracy of $\sim$3.6\% was stated.
We note that a previous independent 
QF-measurement~\cite{CsI-QF-Collar}
provided results consistent with this one
at a $\sim$14\% level.

Folding in published signal efficiencies, 
the efficiency-corrected event rates in 
different $\q2$-bins were derived 
and are depicted in 
Figures~\ref{fig::eff_correct}a\&b
for CsI and Ar, respectively.
These were compared with the complete coherency
conditions in which $\alpha {=} 1 ( \phi {=} 0 )$ and 
equivalently $\Gamma ( \q2 ) {=}  ( \epsilon Z {-} N )^2$ 
is set in Eq~\ref{eq::dsigmadq2}. 
Expected spectra for the completely decoherent 
cases are also displayed.

The systematic uncertainties of this analysis
were taken from the published estimates 
discarding the component due to nuclear form factors.
This corresponds to 11.7\%~\cite{COHERENT-CsI,COHERENT-QF-2020}
and 11.6\%~\cite{COHERENT-Ar}  for CsI and Ar, respectively.
These are represented by the width of 
the complete coherency bands in Figures~\ref{fig::eff_correct}a\&b.
Systematic uncertainties are correlated in $\q2$ in general.
In practice, statistical accuracies 
dominate the uncertainties at the current level of sensitivities,
as shown by comparing the
theory band width with the data error bars
in Figures~\ref{fig::eff_correct}a\&b.
Accordingly, the systematic errors are
{\it assumed} to be uncorrelated in $\q2$.
They are combined bin-wise in quadrature with the statistical errors
to produce the total uncertainties.
Under this error estimation scheme,
the systematic effects contribute to 
$<$4\% and $<$1.5\% of the total uncertainties
over all $\q2$-bins
for CsI and Ar, respectively.

The cross-section reduction fractions
$\xi ( \q2 )$ were evaluated,
from which $\alpha ( \q2)$ and their uncertainties
were extracted using Eq.~\ref{eq::xiVsalpha}.
The various measures which characterize
allowed and excluded intervals at each $\q2$-bin
were then derived with the standard statistics
procedures~\cite{rppstatistics}, 
assuming Gaussian errors.

The allowed 1-$\sigma$ ranges
in $\alpha ( \q2 )$ 
are depicted as stripe-shaded regions in
Figures~\ref{fig::alpha-constraints}a\&b
for CsI and Ar, respectively.
These results are data-driven without 
invoking nuclear physics input.
Measurements and implications 
of every $\q2$-bin are independent of 
and uncorrelated with the others,
and therefore the allowed intervals in adjacent bins
are discontinuous.

The different $\q2$-bins become correlated 
when the nuclear form factors of Eq.~\ref{eq::formfactor} 
are adopted and imposed as theoretical expectations.
The predicted parameter space from
measurements with projected 10\%  uncertainty at $\pm 1 \sigma$-level 
is superimposed as dark-shaded bands,
showing the cases   
with CsI (equivalently, Xe) and Ar at $\DARpi$.
The projected sensitivity for reactor-$\nu$
on Ge is displayed in Figure~\ref{fig::alpha-constraints}c,
showing that measurements with reactor $\nuA$ 
can probe the complete coherency regime.
A bin-wise 10\% uncertainty corresponds to an appropriate
sensitivity target for future experiments. 
It is the scale where systematic effects start to contribute,  
and the measurements can make strong tests 
on the extreme cases of $\alpha {=} 0(1)$
as well as making comparisons with nuclear form factor predictions.

It can be seen that the current data are consistent
with the predictions from Eq.~\ref{eq::formfactor}.
Future measurements with sufficient accuracies
would probe the transitions in QM coherency
according to:
\begin{eqnarray}
{\rm CsI/Xe} ~ & @ & ~ \DARpi:  \nonumber \\
 \alpha & \in &
[ 0.72 , 0.14 ]   {\rm ~~ for ~~}  T \in [6.6,36]~\keVnr  \nonumber \\
{\rm Ar} ~ & @ & ~ \DARpi:   \\ 
 \alpha & \in &
[ 0.88 , 0.51 ]   {\rm ~~ for ~~}  T \in [19,93]~\keVnr \nonumber \\ 
{\rm Ge} ~ & @ & ~ {\rm Reactor}{\it (Projected)}:  \nonumber \\
 \alpha & \in &
[ 1.00 , 0.96 ]   {\rm ~~ for ~~}  T \in [0,1.9]~\keVnr  ~~ ,  \nonumber
\end{eqnarray}
following the $T$-ranges used in Figure~\ref{fig::alpha-constraints}.

The significance in terms of $p$-values~\cite{rppstatistics}
for testing the specific cases 
of $\alpha$=1(0),
equivalently $\phi {=} 0 ( \pi / 2 )$,
with the CsI data are depicted in Figures~\ref{fig::p-values}. 
In particular, the most stringent bounds 
within the stated region-of-interest in Ref.~\cite{COHERENT-CsI} 
in excluding complete QM coherency and decoherency
at 90\% confidence levels with 
specified $p$-values are, respectively,
\begin{eqnarray}
 \alpha  & < & 0.57     \nonumber \\  
 \phi  & < &  0.61 \cdot ( \pi / 2 )   \\  
 p  & = &  0.004    \nonumber          
\end{eqnarray}
at $\q2   {=}   3.1  {\times} 10^{3} ~ {\rm MeV^2}$  
and
\begin{eqnarray}
 \alpha  & > &   0.30     \nonumber \\  
 \phi  & > &  0.80 \cdot ( \pi / 2 )   \\  
 p  & =  & 0.016   ~~  \nonumber
\end{eqnarray}
at $\q2 {=} 2.3  {\times} 10^{3} ~ {\rm MeV^2}$.   
These results verify that both QM superpositions 
among the nucleonic scattering centers
and nuclear many-body effects contribute to the $\nuA$ process.

These diverse ranges of $\alpha$-sensitivity indicate
the complementarity of $\nuA$ measurements among
reactor and $\DARpi$ neutrinos.
Future measurements of solar $\nuA$~\cite{solarnuA} 
with multi-ton detectors would
probe a similar range of $\alpha$ as reactor neutrinos. 
Xenon detectors with scale $\mathcal{O}(100)$ton
would be required to probe the 
weakly-coherent region at $\alpha {<} 0.2$ 
with atmospheric neutrinos.


\input{nuncs-Table2}


\begin{figure*}[t] 
\hspace*{0cm}
{\bf (a)}
\hspace*{8cm}
{\bf (b)}\\
\includegraphics[width=8.2cm]{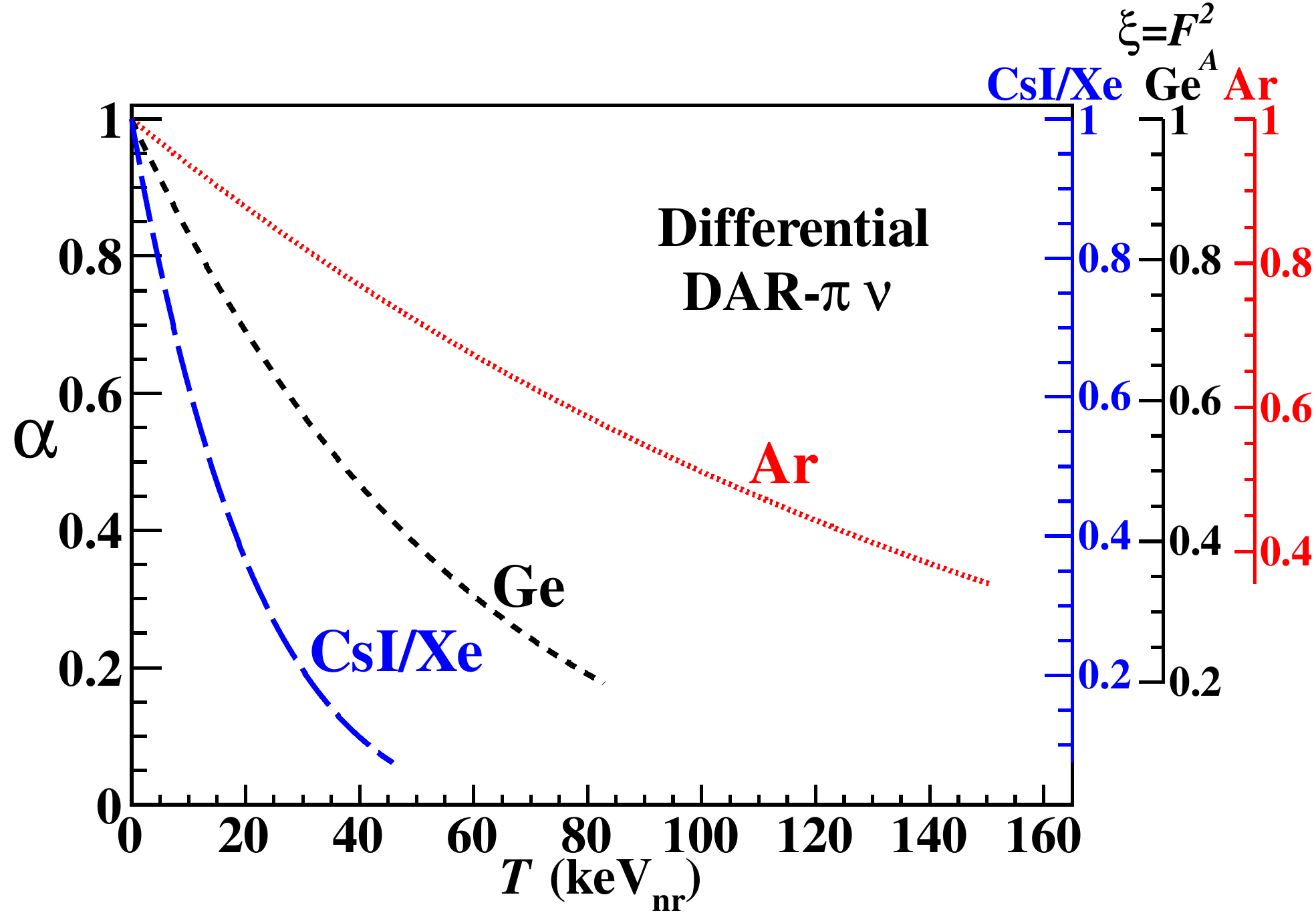}
\includegraphics[width=8.2cm]{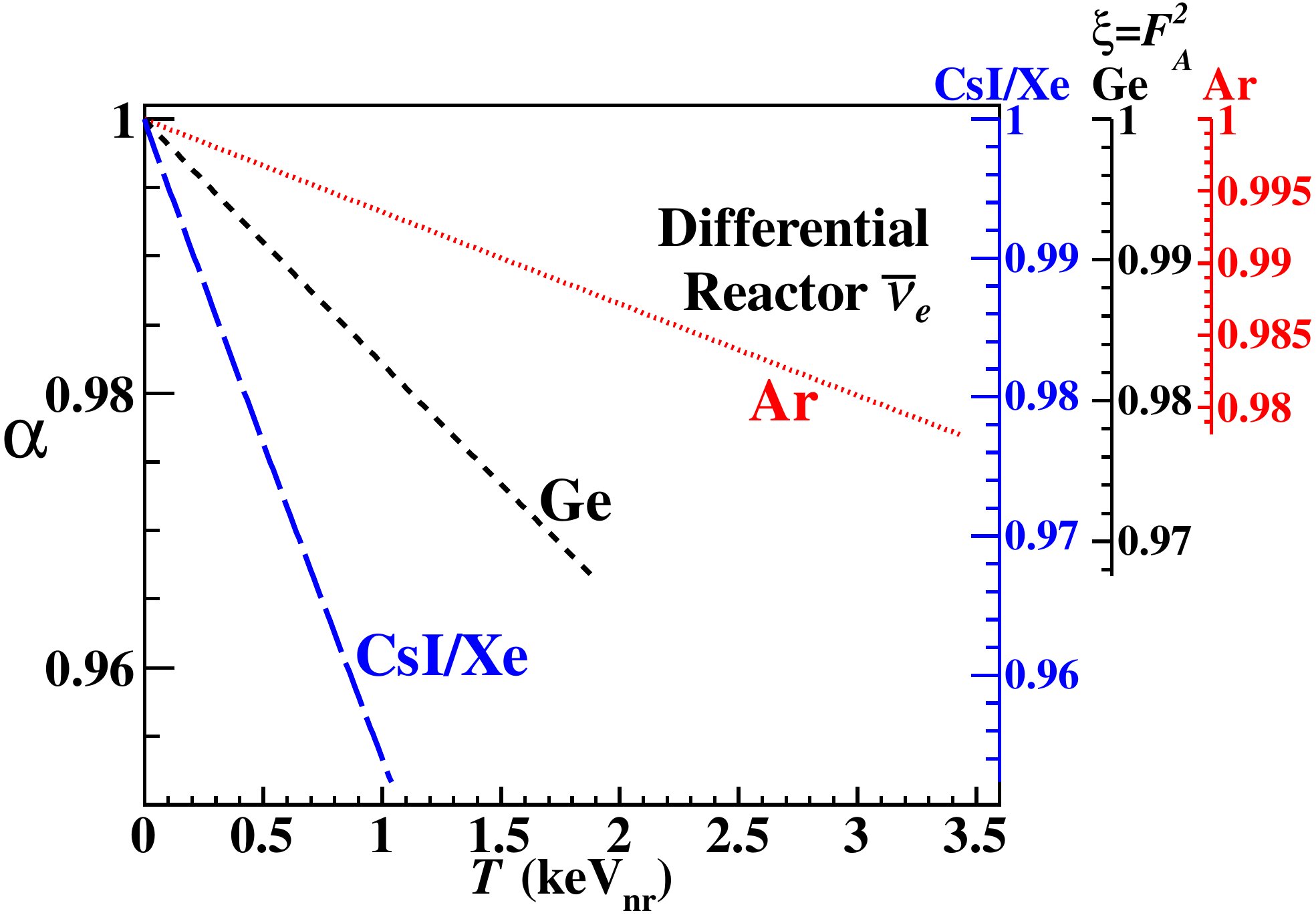}\\
\vspace*{0.2cm}
\hspace*{0cm}
{\bf (c)}
\hspace*{8cm}
{\bf (d)}\\
\includegraphics[width=8.2cm]{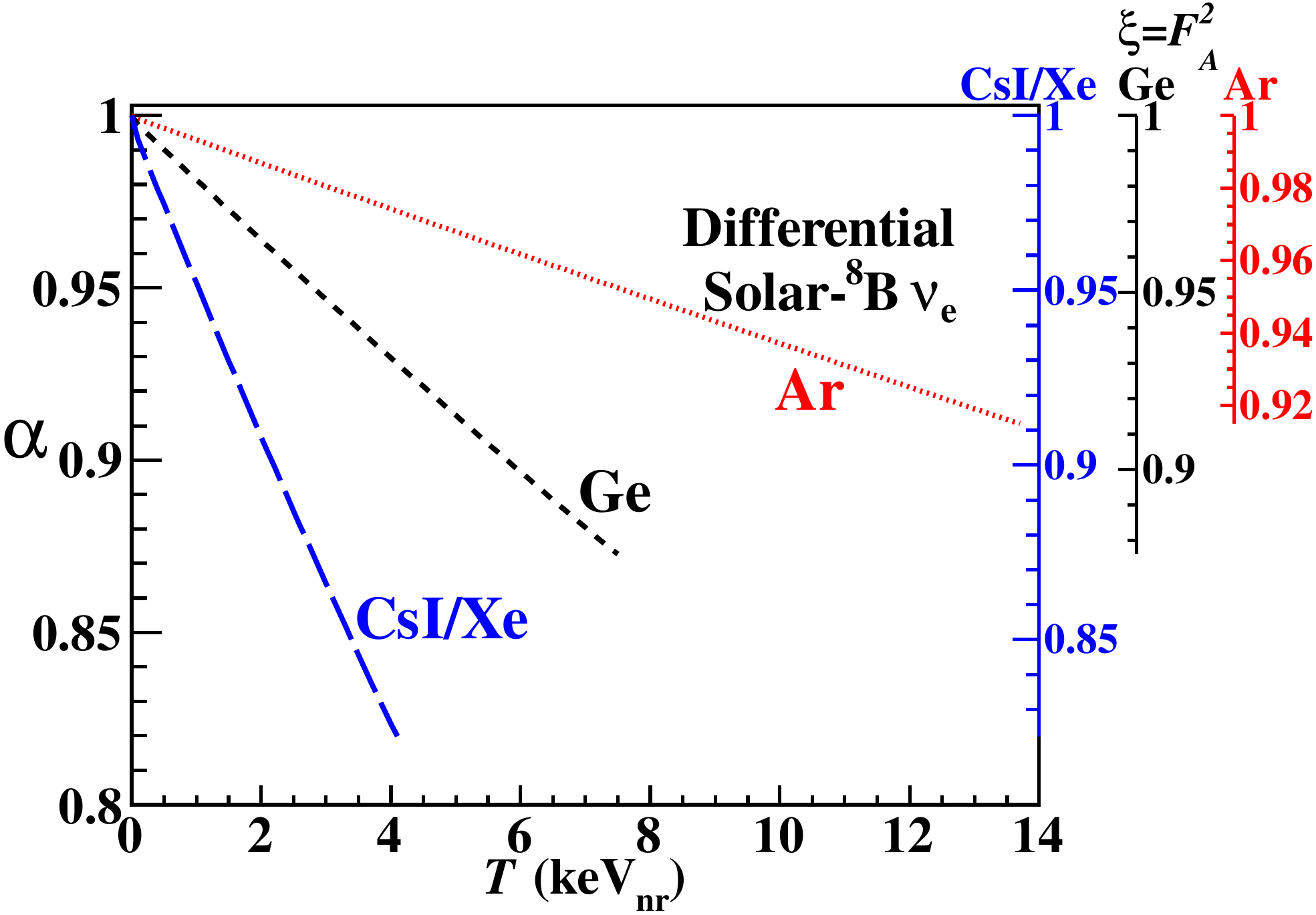}
\includegraphics[width=8.2cm]{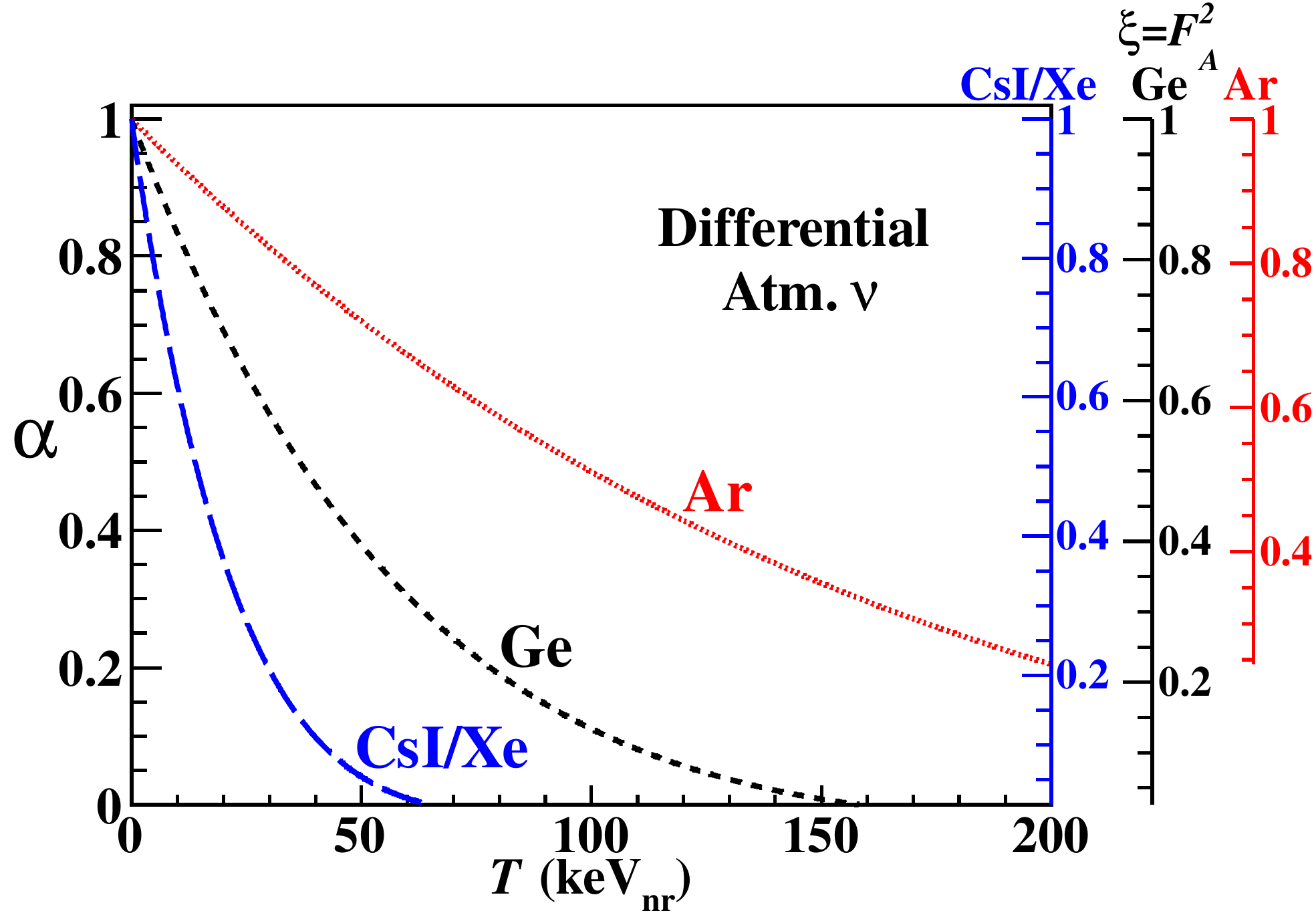}\\
\caption{
The variations of
$\alpha$ and $\xi$(${=} \FF^2$) with $T$
in $\nuA$ on the three selected nuclei for
(a) $\DARpi$,
(b) Reactor,
(c) Solar, and
(d) Atmospheric
neutrinos.
}
\label{fig::TVs3Para}
\end{figure*}


\section{Summary and Prospects}

Neutrino-nucleus elastic scattering
$-$ the interaction $\nuA$ in Eq.~\ref{eq::nuA} $-$
involves two distinct concepts:
elastic kinematics and QM-coherency.
It provides a laboratory to study
QM-superpositions in electroweak interactions.
The QM-coherency aspect should be characterized by
distributions with dependence on $A(Z,N)$ and $\q2$.
Descriptions of coherency as a binary state
or having both concepts bundled together
may have the unintended consequences
of missing the complexities of the process
and suppressing the potential richness of its physics
content.
We formulated a quantitative and universal parametrization
of the QM-coherency~\cite{PRD16}
to facilitate studies of QM-effects in $\nuA$,
under which the fundamental parameter is
the experimentally accessible misalignment phase-angle 
($\phi$, equivalently as $\alpha {\equiv} {\rm cos} \phi$)
between non-identical nucleonic scattering centers.

Current positive measurements on $\nuA$
provide only weak constraints to $\alpha ( \q2 )$ 
and equivalently misalignment phase-angle $\phi ( \q2 )$.
Data with $\mathcal{O}$(10\%) accuracy 
would allow the studies of transitions in QM-coherency 
over a wide range in $\alpha {\in} [0,1]$.

The $\nuA$ process described
in terms of QM-effects ($\GQM$)
is complementary to 
the conventional descriptions 
with nuclear form factors ($\GNP$)
based on the many-body physics in the 
nucleon-nucleus interplay.
Both formulations are related with the directly 
measurable cross-section reduction fraction ($\xi$),
and among themselves,
via Eqs.~\ref{eq::xiVsalpha},\ref{eq::xiVsGNP}\&\ref{eq::FFVsalpha}.

The $\alpha$-parameter quantifies QM-coherency in $\nuA$ 
and adds preciseness to the qualitative discussions.
While the {\it derivation} of the $\alpha$-values depends 
on $A(Z,N)$ and $\q2$ in similar footing as those for $\GNP$,
the {\it interpretation} of measurements
with $\alpha$ in terms of degrees of QM-coherency 
or scattering phase-angle alignment
is universal among different
configurations with varying nuclei and $\q2$.  
This feature allows the quantitative
characterization of the configurations
and is not available with the $\xi$ or $\GNP$ frameworks.
Referring to an example illustrated in Figure~\ref{fig::ParaVsalpha}a,
the four cases ${\rm C_{0,1,2,3}}$
are all within the $\DARpi$ measurable
kinematics domain.
It can be inferred that 
Configuration-${\rm C_0}$ with Ar has 
the same QM-coherency as ${\rm C_2}$ in CsI ($\alpha_0 {=} \alpha_2$)
while having lower and higher levels of coherency
than ${\rm C_1}$ ($ \alpha _0  {<} \alpha _1$)
and ${\rm C_3}$ ($ \alpha _0  {>} \alpha _3$), respectively,
despite the nuclear targets and 
interaction kinematics ($\q2$ and $T$) 
are different and 
the cross-section reduction $ \xi_0 {>} \xi_{1,2,3}$ 
apply in all cases.

New measurements on $\nuA$ 
from a variety of neutrino sources 
and nuclear targets can be expected.
To facilitate comprehensive book-keeping of 
the expanding array of data from diverse configurations,
it would be beneficial to include the $\alpha$-parameter as
a qualifier on QM-coherency to each measurement, 
in the similar spirit as adopting $\q2$ to qualify the interaction kinematics. 

The quantitative description of QM-superpositions 
with the $\alpha$-parameter 
among the scattering amplitudes between individual
nucleons may serve as natural
entry-points to some BSM studies, 
such as those where both coherent and decoherent channels
would contribute to $\nuA$~\cite{EFT}.
The experimentally measurable relation between 
$\GNP$ and $\alpha$ in Figure~\ref{fig::ParaVsalpha}b
describes the transitions in QM-coherency
in terms of the evolution from nuclear to nucleon effects 
in $\nuA$ interaction.
Understanding and applications of these
are possible topics of future research,
but are beyond the scope of the present work.

\section{Acknowledgement}

This work is supported by
the Academia Sinica Principal Investigator Award
AS-IA-106-M02,
Contracts 106-2923-M-001-006-MY5, 107-2119-M-001-028-MY3
and 109-2112-M-259-001
from the Ministry of Science and Technology, Taiwan,
and
2017-ECP2 from
the National Center of Theoretical Sciences, Taiwan.
We are grateful to the Reviewer for bringing to our attention
new quenching factor measurements of Ref.~\cite{COHERENT-QF-2020}
after the initial completion of this work.
This article is dedicated to the memory of Dr. Saime Kerman.

\vspace*{0.3cm}


\begin{figure*}[hbt]
\hspace*{0cm}
{\bf (a)}
\hspace*{7.4cm}
{\bf (b)}\\
\includegraphics[width=7.1cm]{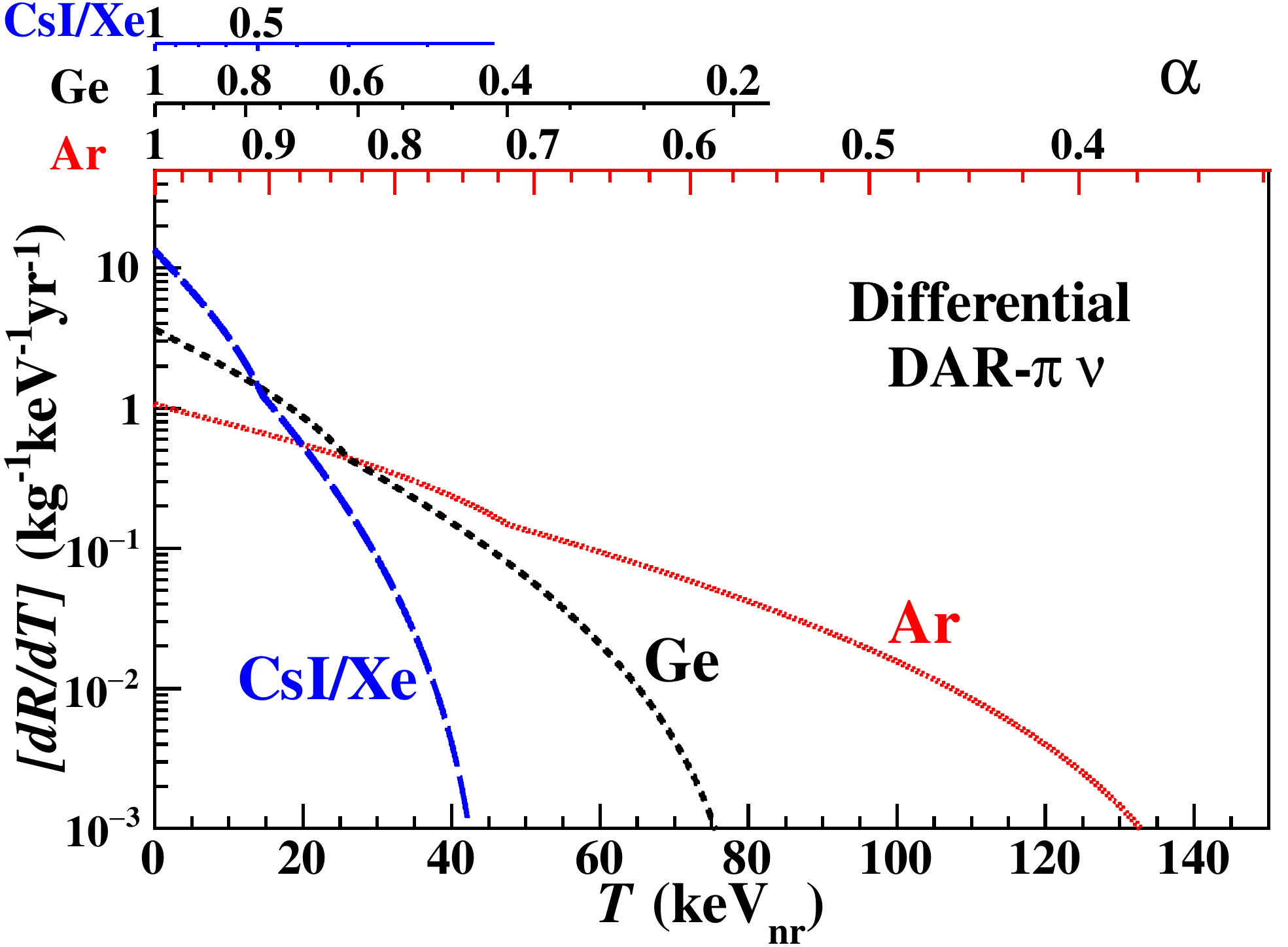}
\hspace*{0.7cm}
\includegraphics[width=7.1cm]{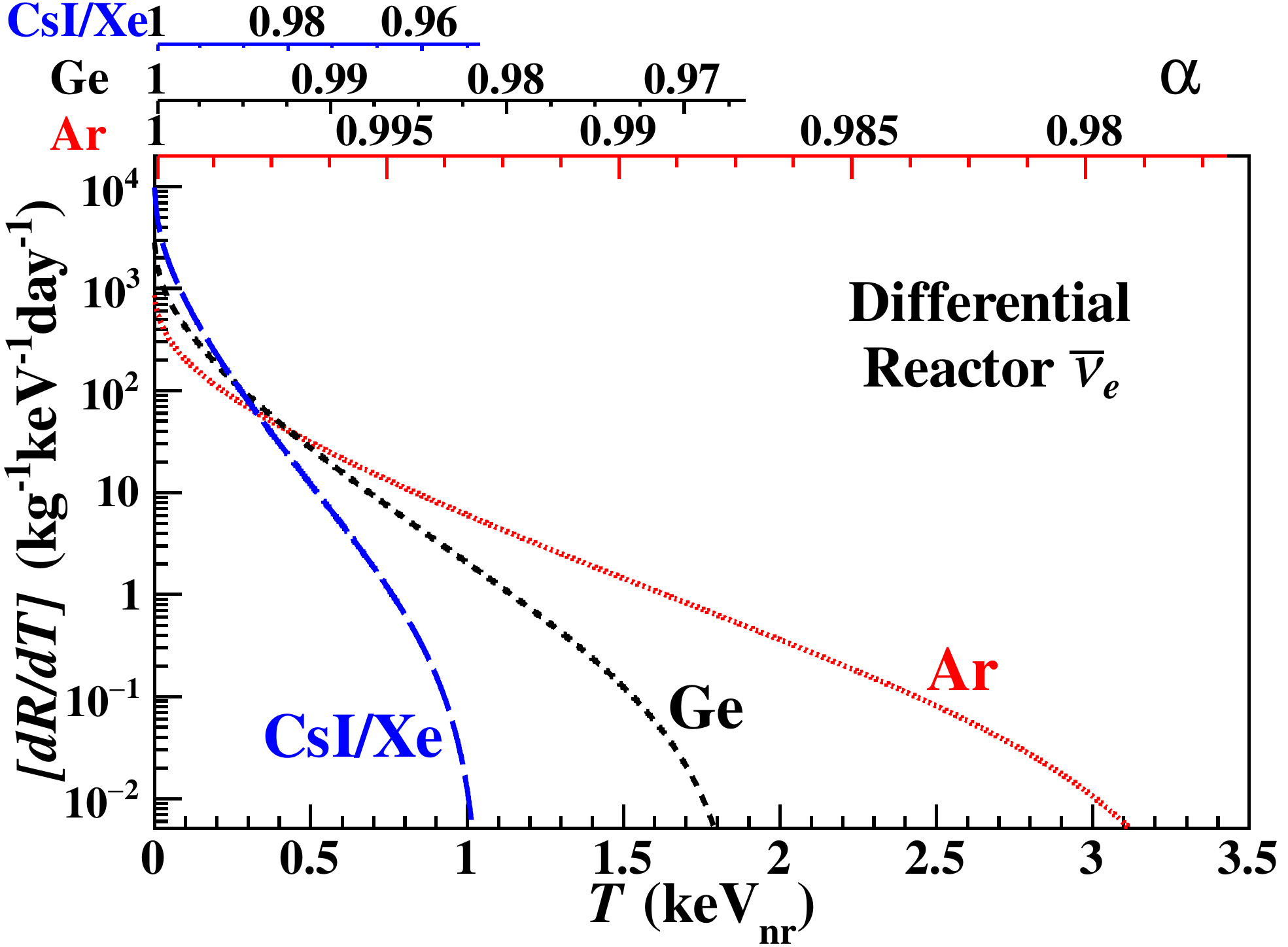}\\
\hspace*{0cm}
{\bf (c)}
\hspace*{7.4cm}
{\bf (d)}\\
\includegraphics[width=7.1cm]{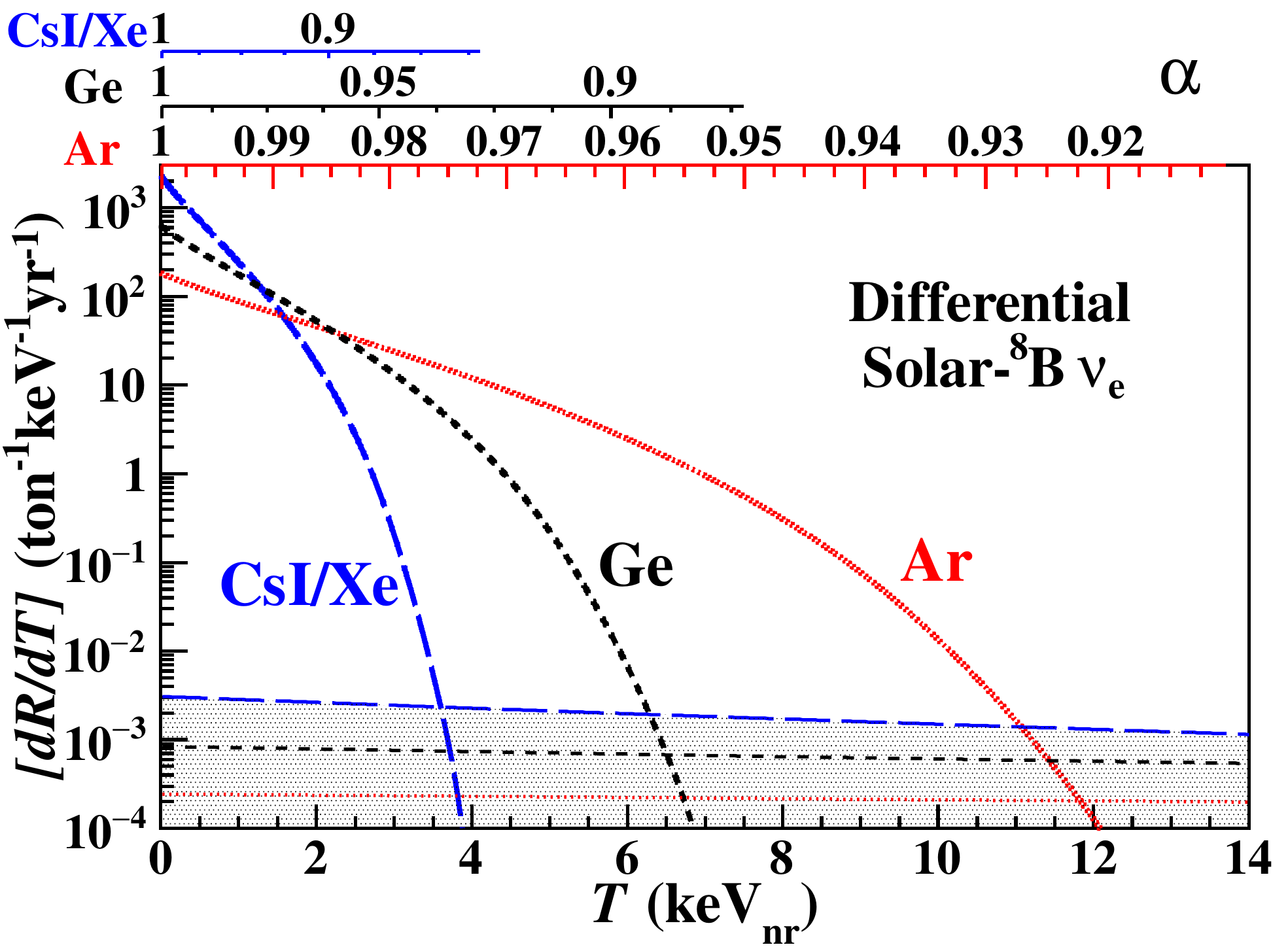}
\hspace*{0.7cm}
\includegraphics[width=7.1cm]{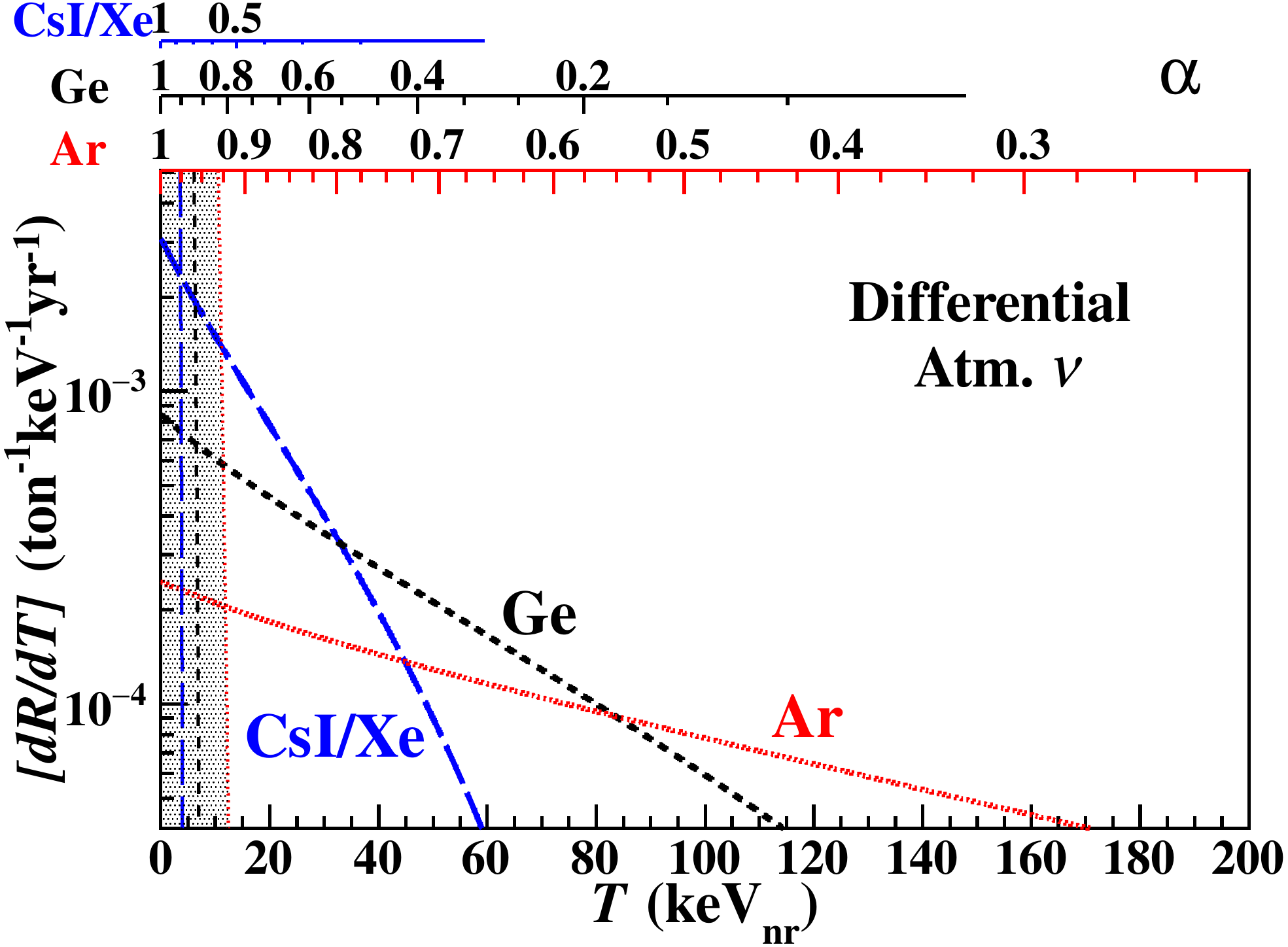}\\
\caption{
Differential event rates $[ dR / dT ]$ of $\nuA$ on the three selected nuclei,
and their correlations with $\alpha$, with
(a) $\DARpi$,
(b) Reactor,
(c) Solar, and
(d) Atmospheric 
neutrinos.
Superimposed as shaded regions 
in (c) and (d) are 
the background rates due to atmospheric and solar neutrinos, respectively.
}
\label{fig::diffcsVsalpha}

\vspace*{0.1cm}

\hspace*{0cm}
{\bf (a)}
\hspace*{7.4cm}
{\bf (b)}\\
\includegraphics[width=7.1cm]{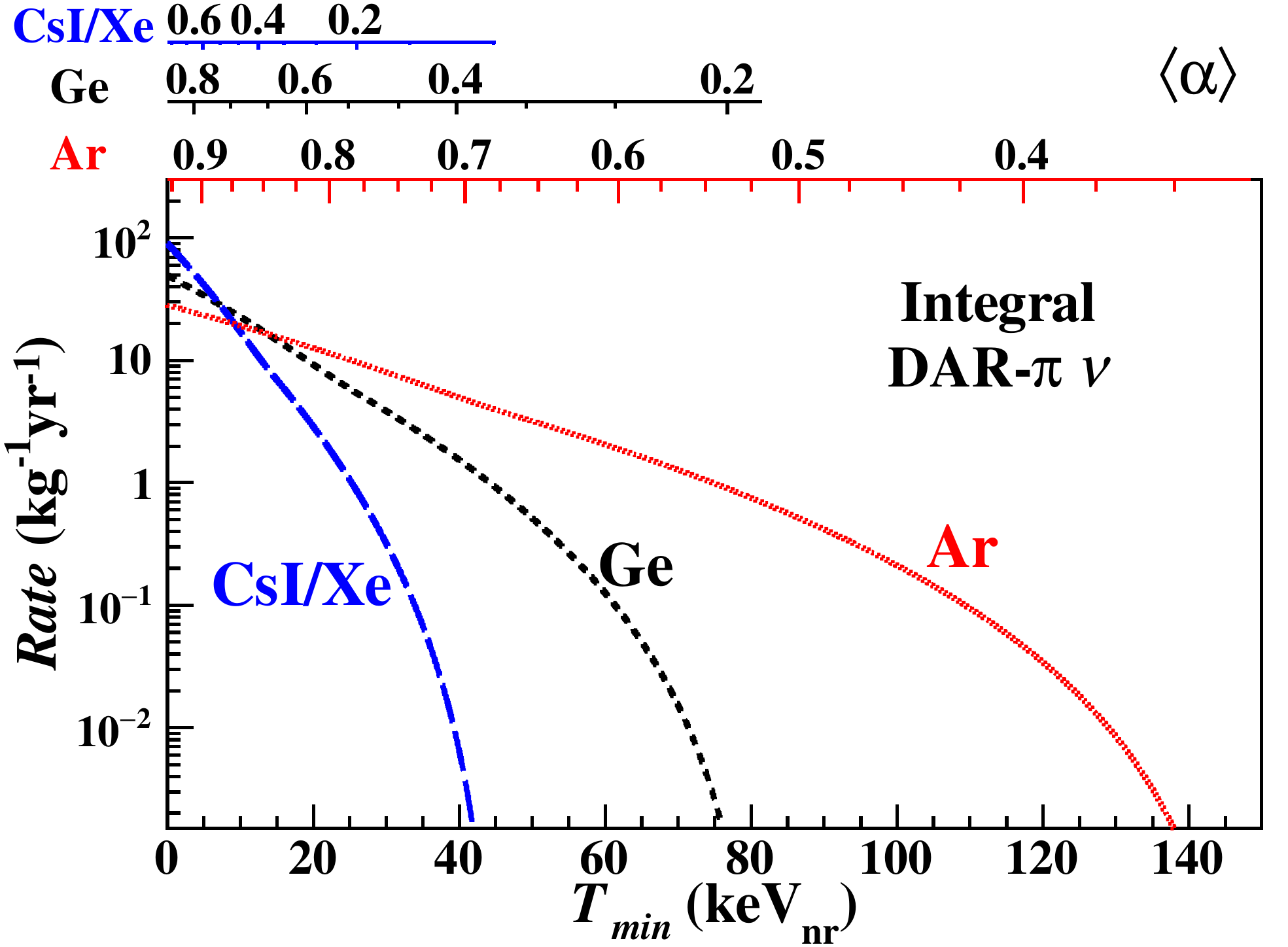}
\hspace*{0.7cm}
\includegraphics[width=7.1cm]{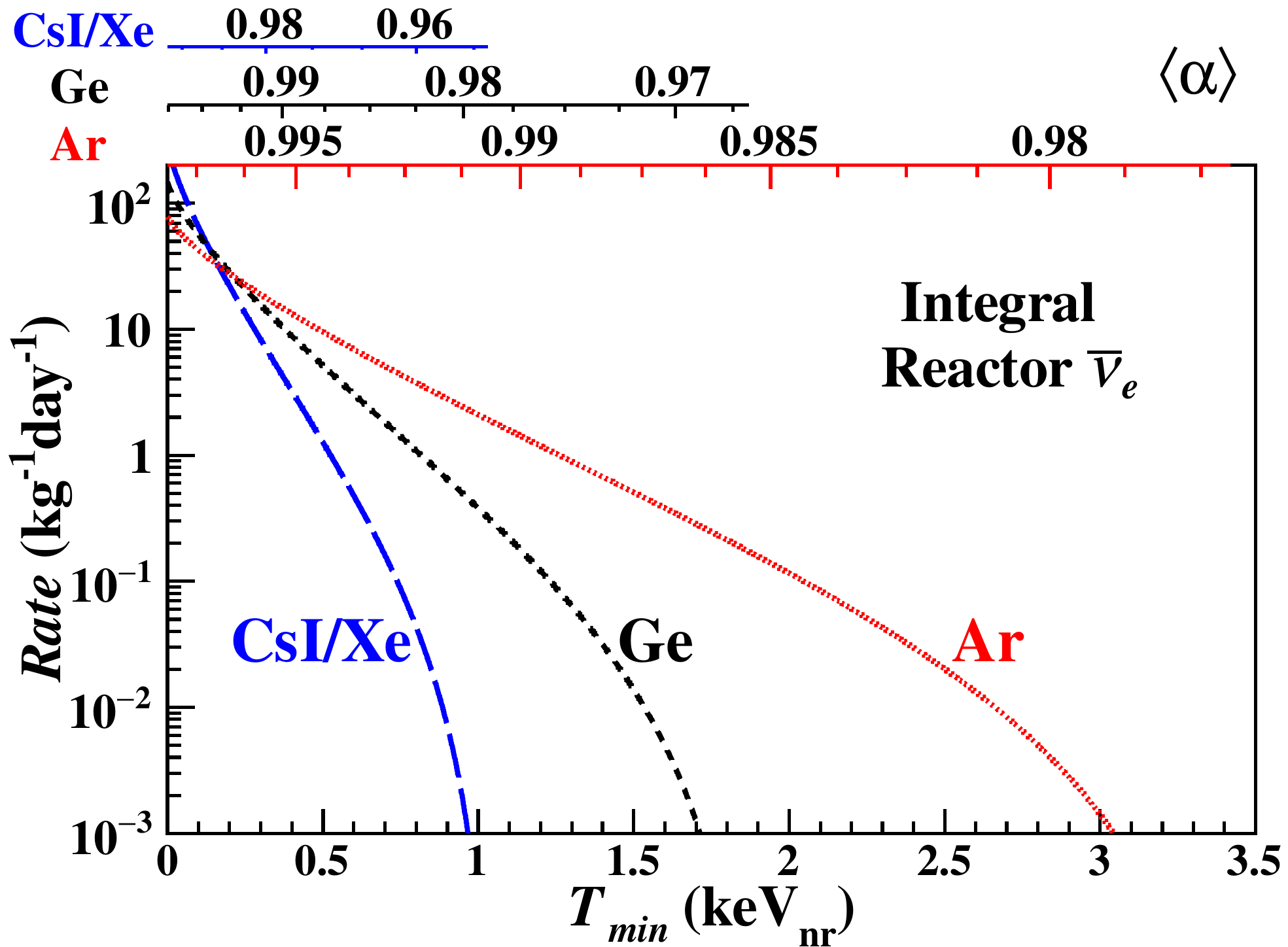}\\
\hspace*{0cm}
{\bf (c)}
\hspace*{7.4cm}
{\bf (d)}\\
\includegraphics[width=7.1cm]{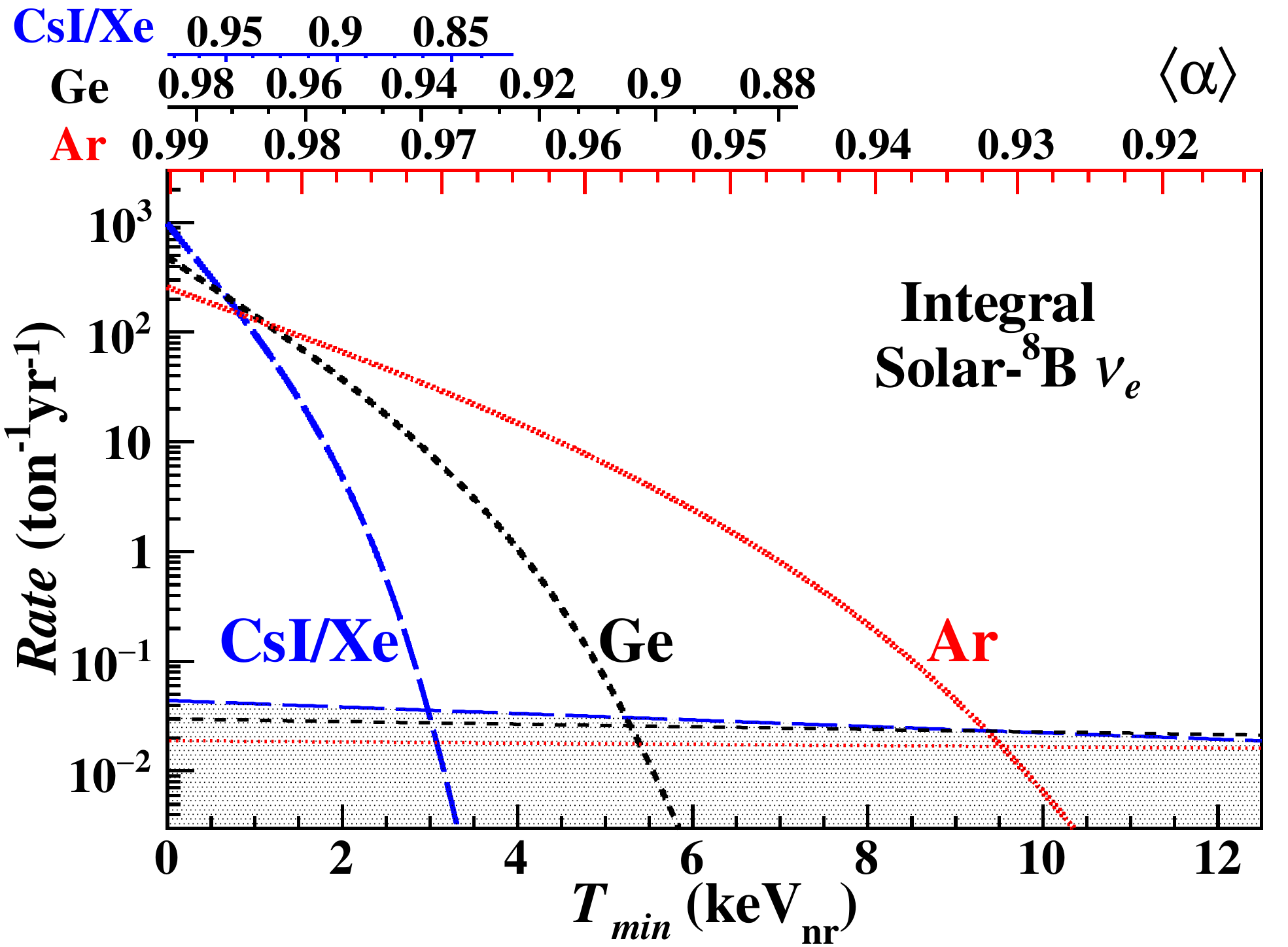}
\hspace*{0.7cm}
\includegraphics[width=7.1cm]{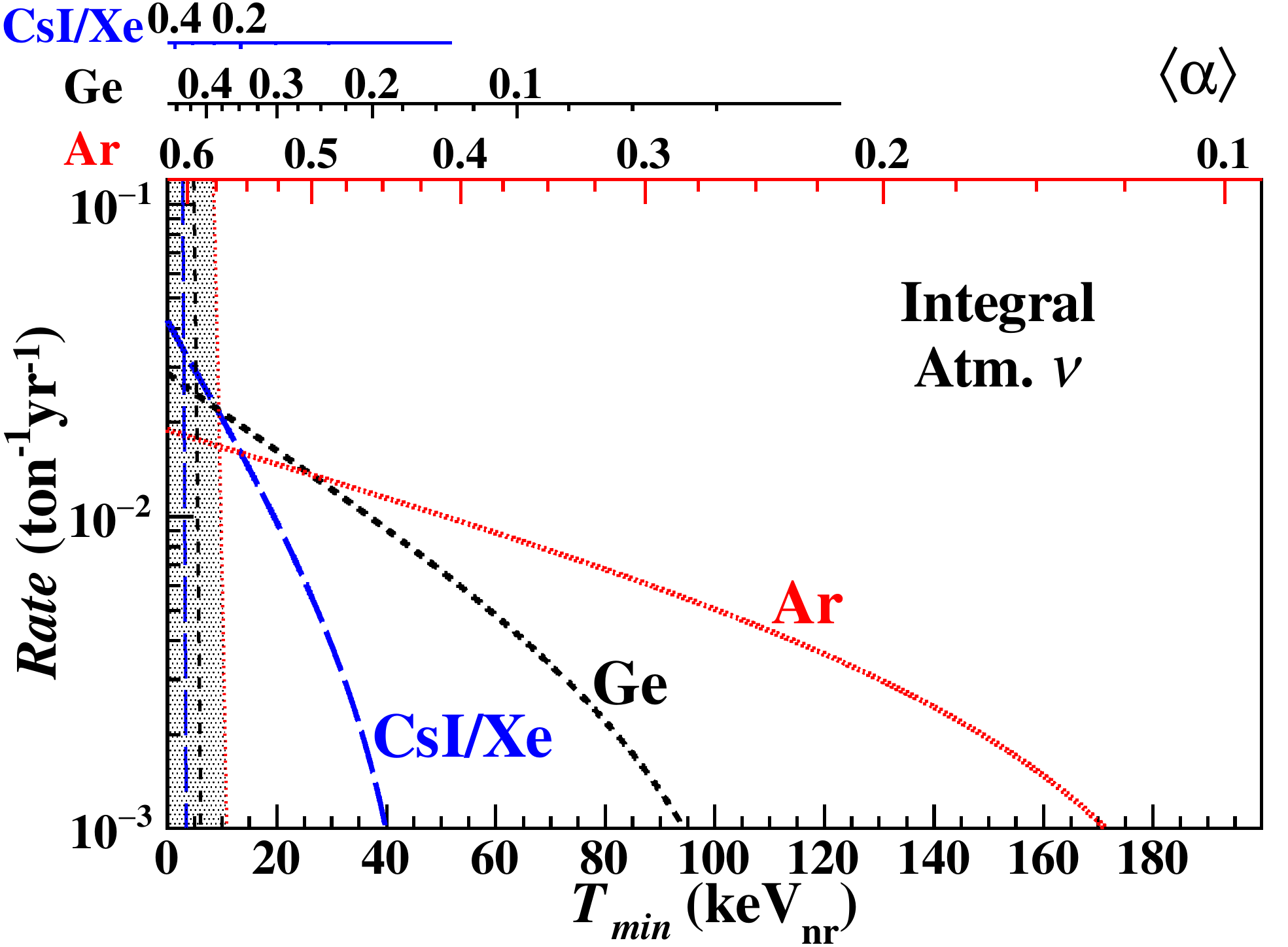}\\
\caption{
Integral event rates of $\nuA$ on the three selected nuclei 
as functions of threshold $\T0$, 
and their correlations with $\avealpha$
for 
(a) $\DARpi$,
(b) Reactor,
(c) Solar, and
(d) Atmospheric 
neutrinos.
Superimposed as shaded regions 
in (c) and (d) are 
the background rates due to atmospheric and solar neutrinos, respectively.
}
\label{fig::intcsVsalpha}
\end{figure*}


\appendix

\section{Measurable Event Rates}
\label{app::rates}

The differential cross-section of Eq.~\ref{eq::dsigmadq2}
on $\q2$ can be translated to 
one as function of the measurable nuclear recoil-energy $T$ 
by
\begin{equation}
\left[ \frac{d \sigma}{d T} \right] _{\nuA} ~   =  ~
 2 M  ~ \left[ \frac{d \sigma}{d \q2} \right] _{\nuA}  ~~  .
\label{eq::dsigmadT}
\end{equation}

The differential spectra $( dR/dT )$ 
convoluted with the neutrino spectrum $\Phi_{\nu} ( \Enu )$
is given by:
\begin{equation}
\left( \dRdT \right)_{\nuA} = 2 M \int
\left[ \left( \frac{d \sigma }{d \q2} \right)_{\nuA} ( T , \Enu )  \right] 
\Phi_{\nu} ( \Enu ) ~ d \Enu  ~~ .
\label{eq::Differential_Rate}
\end{equation}
Integration over $T {\in} [ T_{min} , T_{max} ]$ 
gives the total event rates.

The universality of Figure~\ref{fig::q2-dependence} 
no longer applies when $\q2$ is replaced by $T$.
The variations of $\alpha$, $\FF$ and $\xi$
with $T$ depend on
$\Enu$-distributions and therefore neutrino sources.
The variations are depicted in Figure~\ref{fig::TVs3Para}.


The differential rates
derived from the four sources and three targets 
are displayed in Figure~\ref{fig::diffcsVsalpha}.
The corresponding total rates
are shown in Figure~\ref{fig::intcsVsalpha},
showing their variations with $\T0$ and $\avealpha$.
The values of $\avealpha$ and $\avexi$ as well as the
total event rates at $\T0 {=} 1 (10) ~ \keVnr$
for the various neutrino sources and target nuclei
are summarized in Table~\ref{tab::ave-alpha+xi}.
Evaluation of
these rates are based on standard solar and atmospheric
spectra~\cite{nuspectra}. Reactor $\nuebar$ flux
is taken to be ${\rm 10^{13} ~  cm^{\hyphen 2} s^{\hyphen 1}}$,
while $\DARpi$ per-flavor neutrino flux is
${\rm 3.4 {\times}  10^{14} ~ cm^{\hyphen 2} yr^{\hyphen 1}}$ 
corresponding to
$2 {\times} 10^{23}$~proton-on-target(POT)/year 
at 19.3~m from target~\cite{COHERENT-CsI}. 
There is no high-energy cut-off in $\Enu$ 
for the atmospheric neutrino spectra.
The differential and integral spectra 
of Figures~\ref{fig::diffcsVsalpha}\&\ref{fig::intcsVsalpha}d
are therefore evaluated for $\alpha {\in} [0.01,1.0]$, 
corresponding to $T {<} ( 361 {;} 148 {;} 60 )~ \keVnr$
for (Ar;Ge;Xe).

Typically, measurements of $\nuA$ with reactor and solar neutrinos
require $\mathcal{O}(1)\keVnr$ detector threshold giving expected rates of 
$\mathcal{O}$(1)/kg-day and 
$\mathcal{O}$(1)/ton-yr, respectively.
The corresponding event rates for $\DARpi$ and atmospheric neutrinos are
$\mathcal{O}(10)$/kg-yr and
$\mathcal{O}$(0.01)/ton-yr at a threshold
of $\mathcal{O}(10)\keVnr$, respectively.


At the detection threshold of $1~\keVnr$,
90\% of the elastic scattering events between
Weakly Interacting Massive Particles (WIMPs)-dark matter
of mass 1~TeV with (Ar;Ge;Xe)-target
have recoil energy up to
(99;74;35)~$\keVnr$.
These kinematics ranges correspond to
$\alpha$ as low as (0.49;0.22;0.14)
for $\nuA$ scattering with atmospheric neutrinos,
as indicated in Figure~\ref{fig::TVs3Para}d $-$
far from the complete coherency regime.
Accordingly, the description of ``the neutrino floor
originates from {\it coherent} neutrino-nucleus scattering''
is not applicable
for WIMPs at TeV or higher mass scales.


\end{document}

%% file: nuncs-Table1.tex
\begin{table}
\caption{
Summary of the three formulations which characterize
the many-body physics in $\nuA$, and
the values of the key parameters at the
limiting domains where the 
scattering amplitudes
are either completely in phase (``Coherency'')
or decoupled (``Decoherency'').
}
\label{tab::QMcoherency}
\centering
\begin{center}
\begin{tabular}{|ccc|}
\hline 
\multirow{2}{*}{Conditions ~~~~} & Complete  & Complete  \\
  &  ~~~~~~ Coherency ~~~~~~ &  ~~~~ Decoherency ~~~~ \\ \hline \hline
\multirow{3}{*}{$\q2$} & 
\multirow{3}{*}{$\rightarrow 0$} & 
\multirow{2}{*}{$\scaleto{ \gtrsim [ \frac{\pi}{R} ]^2}{14pt}$} \\
& & \\
& & with $A$-Dependence \\
& & \\
\multicolumn{3}{|l|}{(I) $\GNP ( \q2 ) {=}
 \left[ \varepsilon Z F_Z ( \q2 ) {-} N F_N ( \q2 ) \right] ^2 $} \\
 ~~ $F_Z ( \q2 )$ & 1 & $-$ \\
 ~~ $F_N ( \q2 )$ & 1 & $-$ \\
 $\GNP ( \q2 )$ & $ ( \varepsilon  Z - N ) ^2 $ &  
$ ( \varepsilon^2  Z + N )  $   \\ 
& & \\
\multicolumn{3}{|l|}{(II) $\GQM ( \q2 ) {=}
 ( \varepsilon Z {-} N ) ^2 \alpha ( \q2 )   {+}  
( \varepsilon ^2  Z {+} N )  
\left[ 1 {-} \alpha ( \q2 ) \right]$} \\ 
 $\phi ( \q2 )$ & 0 & $\pi$/2\\
 $\alpha ( \q2 )$ & 1 & 0 \\
& & \\
\multicolumn{3}{|l|}{(III) $\GDATA ( \q2 ) {=} 
( \varepsilon Z {-} N ) ^2  \xi ( \q2 ) $} \\ 
\multirow{2}{*}{$\xi ( \q2 )$} & 
\multirow{2}{*}{1} & 
\multirow{2}{*}{$\left[\scaleto{   \frac{ ( \varepsilon ^2  Z + N ) }
{ ( \varepsilon  Z - N  ) ^2 }}{18pt} \right]$} \\
& & \\
& & \\
\multirow{2}{*}{$\left[ \scaleto{ \frac{ d \sigma}{d \q2}}{18pt}  \right] ( \q2 )$} & 
\multirow{2}{*}{${\propto} ( \varepsilon  Z - N ) ^2 $} &  
\multirow{2}{*}{${\propto} ( \varepsilon^2  Z + N )  $}   \\ 
& & \\
\hline
\end{tabular}
\end{center}
\end{table}

%% file: nuncs-Table2.tex

\begin{table*}
\caption{
Averaged $ [ \avealpha ; \avexi ] $ 
and total event rates in ${\rm kg^{\hyphen 1} day^{\hyphen 1} }$
for the target nuclei at a threshold of 1 and 10~$\keVnr$ and
for different $\nu$-sources.
Reactor and $\DARpi$ neutrino fluxes are taken to be
${\rm 10^{13} ~  cm^{\hyphen 2} s^{\hyphen 1}}$,
while $\DARpi$ neutrino flux is
${\rm 3.4 {\times} 10^{14} ~  cm^{\hyphen 2} yr^{\hyphen 1} / flavor}$ 
at 19.3~m from target at beam intensity 
${\rm 2 {\times} 10^{23} ~ {\rm POT} ~ yr^{\hyphen 1}}$. 
Rates due to atmospheric neutrinos are from the integration 
of $\q2$-ranges corresponding to $\alpha {\in} [0.01,1.0]$.
}
\begin{center}
  \begin{tabular}{|c||c|c|c|c|}
\hline 
 \multirow{4}{*}{Detector Target} 
& \multicolumn{4}{c|}{ $\nu$-Sources}\\ 
& \multicolumn{1}{c}{ $\DARpi$ } & \multicolumn{1}{c}{ Reactor } & 
\multicolumn{1}{c}{ Solar } & \multicolumn{1}{c|}{ Atmospheric }   \\ \cline{2-5}  
& \multicolumn{2}{c|}{  $\left[ ~ \avealpha ~ ; ~ \avexi ~ \right]$  }  
& \multicolumn{2}{c|}{  $\left[ ~ \avealpha ~ ; ~ \avexi ~ \right]$  }  \\
& \multicolumn{2}{c|}{ Total Event Rates ${\rm ( kg^{\hyphen 1} yr^{\hyphen 1} )}$} 
& \multicolumn{2}{c|}{ Total Event Rates ${\rm ( ton^{\hyphen 1} yr^{\hyphen 1} )}$}   \\ \hline \hline


 & \multicolumn{4}{c|}{ \multirow{2}{*}{
Detector Threshold = $1 ~ \keVnr$ } } \\ 
& \multicolumn{4}{c|}{}  \\ \cline{2-5}
    \multirow{2}{*}{Ar}  & ~~ [ 0.92 ; 0.93 ] ~~  & ~~  [ 0.99 ; 0.99 ] ~~  
& ~~ [ 0.98 ; 0.98 ] ~~    & ~~ [ 0.61 ; 0.63 ] ~~   \\
        &  27.2 &   766 &  130  &  0.019 \\
    \multirow{2}{*}{Ge}  & [ 0.84 ; 0.84 ]    & [ 0.98 ; 0.98 ]   
& [ 0.97 ; 0.97 ]   & [ 0.46 ; 0.47 ]   \\
        & 46.1  & 138 &  140  &  0.028 \\
    \multirow{2}{*}{Xe}  & [ 0.72 ; 0.72 ]   & [ 0.95 ; 0.95 ]   
& [ 0.94 ; 0.94 ]    & [ 0.41 ; 0.42 ]    \\
        & 77.8  &  0.07 &  95.7  &  0.039 \\ \hline


 & \multicolumn{4}{c|}{ \multirow{2}{*}{
Detector Threshold = $10 ~ \keVnr$ } } \\ 
& \multicolumn{4}{c|}{}  \\ \cline{2-5}

    \multirow{2}{*}{Ar}  & [ 0.87 ; 0.87 ]  & 
\multirow{2}{*}{N/A}     
& \multirow{2}{*}{N/A} & [ 0.57 ; 0.59 ]   \\
      &  19.2  &     &  &  0.017 \\
    \multirow{2}{*}{Ge}  & [ 0.72 ; 0.72 ]   & \multirow{2}{*}{N/A}  
& \multirow{2}{*}{N/A} & [ 0.37 ; 0.39 ]  \\
        & 22.6  &   &  &  0.022 \\
    \multirow{2}{*}{Xe}  & [ 0.46 ; 0.47 ]   & \multirow{2}{*}{N/A}   
& \multirow{2}{*}{N/A} & [ 0.24 ; 0.25 ]   \\ 
      & 17.1  &   &  &  0.021 \\ 
\hline
  \end{tabular}
\end{center}
\label{tab::ave-alpha+xi}
\end{table*}

%% file: nuncs-arXiv.bbl
\begin{thebibliography}{99}
\bibitem{footnote1}
The $\alpha$ and $\xi$ results of Ref.~\cite{PRD16}
are mean values equivalent to $\avealpha$ and $\avexi$, respectively, 
in this article.
\end{thebibliography}
